\documentclass{mnras}

\usepackage{newtxtext,newtxmath}
\usepackage[T1]{fontenc}
\usepackage{comment}

\DeclareRobustCommand{\VAN}[3]{#2}
\let\VANthebibliography\thebibliography
\def\thebibliography{\DeclareRobustCommand{\VAN}[3]{##3}\VANthebibliography}

\usepackage{graphicx}	
\graphicspath{{./figures/}}
\usepackage{float}
\let\bibhang\relax
\usepackage{natbib}
\usepackage{amsmath}	

\title[Binary Evolution in 7 Magnetic DA White Dwarfs]{Signs of Binary Evolution in 7 Magnetic DA White Dwarfs}
\author[Moss et al.]{
Adam Moss,$^{1}$
Mukremin Kilic,$^{1}$
P. Bergeron,$^{2}$
Megan Firgard,$^{1}$
Warren Brown$^{3}$
\\
$^{1}$Homer L. Dodge Department of Physics and Astronomy, University of Oklahoma, 440 W. Brooks St., Norman, OK 73019, USA\\
$^{2}$Département de Physique, Université de Montréal, C.P. 6128, Succ. Centre-Ville, Montréal, Québec H3C 3J7, Canada\\
$^{3}$Smithsonian Astrophysical Observatory, 60 Garden Street, Cambridge, MA 02138, USA
}

\date{Accepted 2023 June 15}
\pubyear{2023}

\begin{document}
\label{firstpage}
\pagerange{\pageref{firstpage}--\pageref{lastpage}}
\maketitle

\begin{abstract}
We present our findings on the spectral analysis of seven magnetic white dwarfs that were presumed to be double degenerates. We obtained time-resolved spectroscopy at the Gemini Observatory to look for evidence of binarity or fast rotation. We find three of our targets have rotation periods of less than an hour based on the shifting positions of the Zeeman-split H$\alpha$ components: 13, 35, and 39 min, and we find one more target with a $\sim$hour long period that is currently unconstrained. We use offset dipole models to determine the inclination, magnetic field strength, and dipole offset of each target. The average surface field strengths of our fast rotators vary by 1$-$2 MG between different spectra. In all cases, the observed absorption features are too shallow compared to our models. This could be due to extra flux from a companion for our three low-mass targets, but the majority of our sample likely requires an inhomogeneous surface composition. Including an additional magnetic white dwarf with similar properties presented in the literature, we find that 5 of the 8 targets in this sample show field variations on minute/hour timescales. A crystallization driven dynamo can potentially explain the magnetic fields in three of our targets with masses above $0.7~M_{\odot}$, but another mechanism is still needed to explain their rapid rotation. We suggest that rapid rotation or low-masses point to binary evolution as the likely source of magnetism in 7 of these 8 targets.
\end{abstract}

\section{Introduction}
The origin of magnetic fields in white dwarfs (WDs) remains a mystery. Proposed formation channels generally fall into two categories: a fossil origin or an evolutionary origin. In the fossil case, the collapse from a giant star to a WD results in a strong magnetic field due to the conservation of magnetic flux \citep{Tout04}. Therefore, WDs formed from progenitor Ap/Bp stars are an appealing solution given their stronger than average magnetic fields. However, this formation channel by itself cannot account for all of the magnetic WDs due to the lower than required incidence of Ap/Bp stars \citep{Wick05}.

Multiple evolutionary scenarios could further produce magnetic WDs. The onset of core crystallization could trigger a dynamo effect \citep{Isern17,Schreiber21}. However this process can explain some, but not all, of the isolated white dwarfs with MG fields
\citep{Ginzburg22}. Binary mergers are a strong candidate given the lack of magnetic WDs in detached binaries with main sequence stars \citep{Liebert05}. \citet{Liebert15} confirmed their findings and did not find any magnetic WDs in WD-MS binaries regardless of separation besides cataclysmic variables. \citet{Landstreet20} found five binary systems with a magnetic WD and a non-degenerate companion, however four of these were too widely separated to undergo Roche-lobe overflow in the future. \citet{Parsons21} found six magnetic WDs in detached binaries with low-mass stellar companions, but determined they are temporarily detached cataclysmic variables as opposed to pre-cataclysmic binaries. The higher than average mass of magnetic WDs, $\sim$ 0.8$M_\odot$, further supports the
binary merger channel as a likely source of (at least some of the) strongly magnetic WDs \citep{Ferrario15a,Briggs15,Briggs18}.

Modeling the magnetic fields in WDs presents its own challenges, with some objects needing "simple" models while others requiring more complicated ones such as an offset dipole or quadrupole. \citet{Rolland15} analyzed 16 cool magnetic DA WDs and were able to successfully fit the spectra of 6 using an offset dipole model. The remaining 10 resulted in discrepancies between the photometric and spectroscopic temperatures. The explanation at the time was that each WD was in an unresolved binary system with a DC WD, effectively diluting the H$\alpha$ line profile. There are several examples of magnetic WDs in double degenerate binaries \citep[see][for a list]{Kawka17}. However, such a large incidence (10 out of 16) of magnetic WDs in double degenerate systems would be inconsistent with population synthesis calculations \citep{Briggs15} and the rate of Type Ia supernovae (e.g., \citep{Ruiter20}).

\citet{Kilic19} conducted follow up analysis on one of these targets, G183-35. They detected significant variations in the Zeeman-split H$\alpha$ wavelength positions over the span of 2.9 hours. This target shows five split components, with the inner and outer lines appearing and disappearing alternatively over the observations, indicating that the variations are due to rotation of the WD. Hence, Kilic et al. hypothesized that G183-35 is a single WD with a magnetic axis offset from the rotation axis, resulting in a different magnetic field distribution as the WD rotates. An inhomogeneous H/He atmosphere was further proposed to resolve the issue of diluting the unusually shallow H$\alpha$ line profile to match the observations. 

\begin{figure}
    \centering
    \includegraphics[width=3.4in, clip=true, trim=0in 0.1in 0.5in 0.5in]{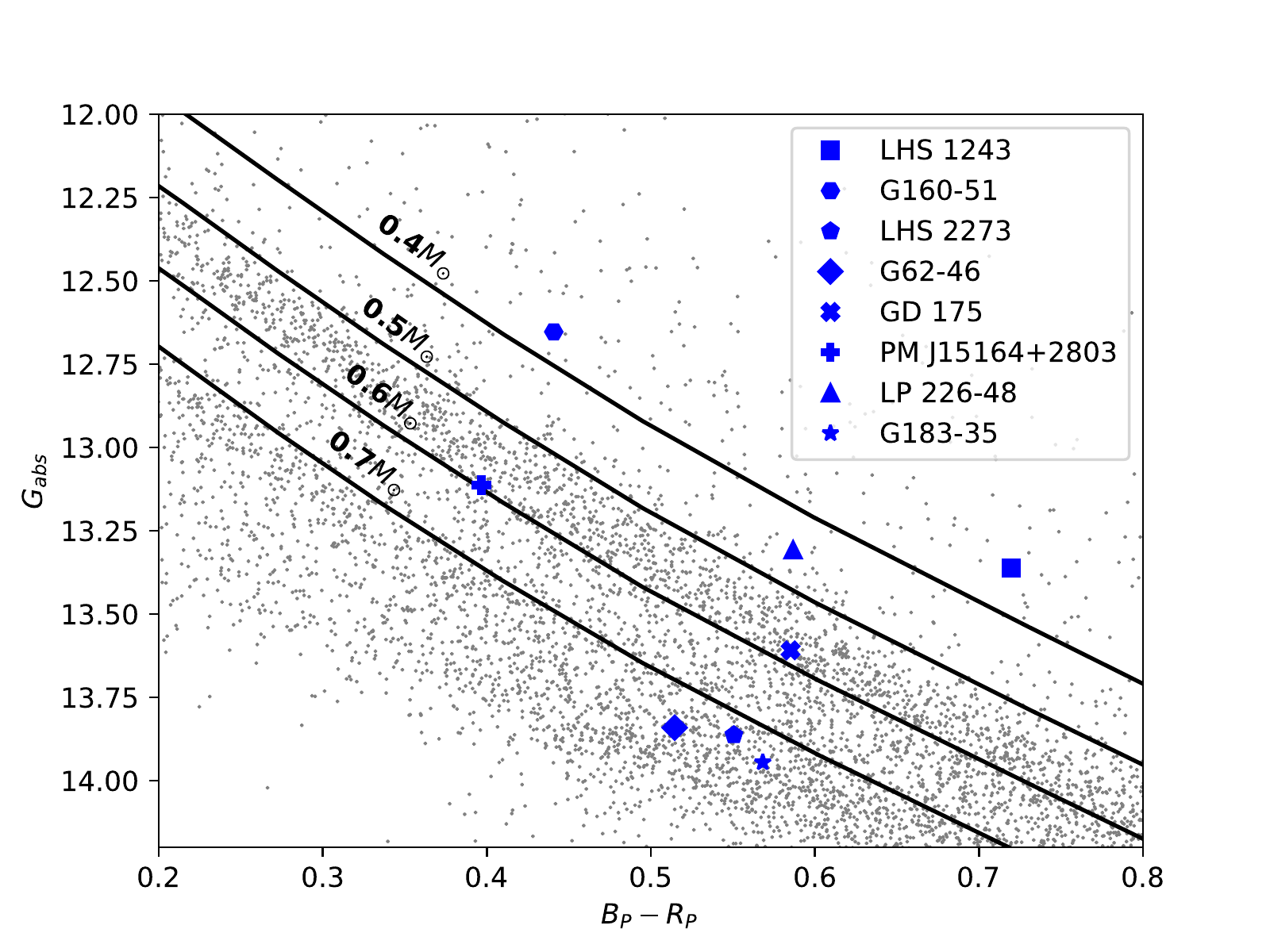}
    \caption[Caption]{Our sample of eight white dwarfs (including G183-35 from \citealt{Kilic19}) in the Gaia color-magnitude diagram. Gray points are white dwarfs within 100 pc from the \cite{Gentile21} catalog. Black lines are theoretical cooling models for white dwarfs of various masses. \footnotemark[1]}
    \label{fig1}
\end{figure}
\footnotetext[1]{See http://www.astro.umontreal.ca/$\sim$bergeron/CoolingModels.}

\begin{table*}
\begin{center}
\caption{\label{tab1}Summary of observations for each target. The format for the exposures is (number of exposures) x (exposure duration in min).}
\begin{tabular}{|c|c|c|c|c|c|} 
 \hline
Object & Gaia DR3 ID & RA(J2000) & Dec.(J2000) & Short Exp. & Long Exp.\\
 \hline
 LHS 1243 & 4983839647522981504 & 01:24:03.9 & $-$42:40:38.4 & $16\times1.33$ min & $27\times5$ min \\ 
 G160-51 & 3189621320226676736 & 04:12:26.3 & $-$11:17:47.3 & None & $27\times5$ min \\
 LHS 2273 & 3882611201058534400 & 10:29:07.5 & +11:27:19.3 & None & $25\times4.1$ min \\ 
 G62-46 & 3711214067185666560 & 13:32:50.7 & +01:17:06.3 & $14\times2$ min & $15\times5$ min \\
 GD 175 & 6332763530870415488 & 15:05:49.3 & $-$07:14:40.9 & $14\times2$ min & $55\times3$ min \\
 PM J15164+2803 & 1271649969930799872 & 15:16:25.1 & +28:03:20.9 & $14\times2$ min & $15\times5$ min \\ 
 LP 226-48 & 1341543072245722752 & 17:14:50.8 & +39:18:37.4 & $14\times2$ min & $17\times4$ min \\
 \hline
\end{tabular}
\end{center}
\end{table*}

Given the complexities of G183-35, analysis of the remaining 9 targets in the sample is necessary to determine if they also show variations in their spectra and can be classified as isolated objects. We were awarded time at the Gemini Observatory to obtain time-resolved spectroscopy for all 9 targets and were able to observe 7 of them. 

Figure \ref{fig1} shows the location of these 7 targets plus
G183-35 in a color-magnitude diagram of the 100 pc WD sample. Interestingly, several of our targets seem to be over-luminous, which is indicative of a low-mass or excess flux from a companion.
Here we present our findings for each system. Section 2 details our observations, and in Section 3 we present our results for each target. We discuss the nature of each object including constraints on their rotation periods and crystallization states, and conclude in Section 4. 

\section{Observations}

We obtained time-resolved spectroscopy using the 8m Gemini North and South telescopes with the Gemini Multi-Object Spectograph (GMOS) as part of the following programs: GS-2019B-FT-107, GS-2020A-Q-311, GS-2021A-Q-136, GS-2021A-Q-321, GN-2020A-Q-116, GN-2021A-Q-135, and GN-2021A-Q-318. Given the typical short rotation periods of magnetic WDs, our programs were designed to sample periods on the order of minutes and hours. Our observations are summarized in Table \ref{tab1}. We used the R831 grating and a 0.5" slit, providing wavelength coverage from 5350 Å to 7710 Å and a resolution of 0.376 Å per pixel. A comparison lamp exposure was taken after every $\sim$15 minutes of science exposures for each target. We used the IRAF Gemini gmos package to reduce these data.  

For LHS 1243, our initial sampling was not fine enough to constrain the rotational period. We obtained additional
spectra for LHS 1243 at the 6.5m Magellan telescope with the MagE spectrograph. We used the $0.85\arcsec$ slit, providing
wavelength coverage from about 3400 \AA\ to 7000 \AA\ with a resolving power of R = 4800. Given the short rotational period, we obtained $16\times80$ s exposures to better sample the period.

\section{Results}
Of our 7 targets, we are able to constrain the rotational period for 3 targets (LHS 1243, LHS 2273, PM J15164+2803) based on the periodic variations in their line centers. We generate Lomb-Scargle periodograms using the orbital fit code MPRVFIT \citep{DeLee13} for these 3 targets. One target (G62-46) shows clear variations but we were unable to observe it long enough to constrain its period. The remaining targets do not show variations in their trailed spectra, likely because their rotation periods are longer than a few hours. The analysis of each target is presented in the following subsections.

Before fitting the spectra with an offset dipole model, we construct atmosphere models for each target based on their $T_{\rm eff}$ and $\log{g}$ as  determined from \citet{Caron23}, who used Gaia DR3 parallaxes, Pan-STARRS $grizy$, and near-IR JHK photometry. The physical parameters for each target are given in Table \ref{tab2}.

\begin{table*}
\begin{center}
\caption{\label{tab2} Physical parameters of all 8 white dwarfs in our sample. The effective temperature and masses are derived from the photometric fits from \citet{Caron23}. For the fast rotators, we list the range of magnetic field strengths found in our fits.}
\begin{tabular}{|c|c|c|c|c|c|c|c|} 
 \hline
 WD ID & $T_{\rm eff}$ [K] & Mass [M$_{\odot}$] & Rotation Period [hr] & $B_d$ [MG] & $v_T$ [km/s] & Cooling Age [Gyr] & DC Offset\\
 \hline
 LHS 1243 & 6400 & 0.430 & 0.216 & 10$-$10.6 & 53 & 1.4 & 1.5\\ 
 G160-51 & 7450 & 0.385 & N/A & 3.2 & 47 & 0.8 & 0.1\\
 LHS 2273 & 7041 & 0.806 & 0.648 & 16.0$-$17.2 & 78 & 3.1 & 1.5\\ 
 G62-46 & 7204 & 0.829 & $\sim$hr & 6.3$-$7.7 & 54 & 3.1 & 1.5\\
 GD 175 & 6889 & 0.647 & N/A & 3.2 & 26 & 1.9 & 1.1\\
 PM J15164+2803 & 8073 & 0.682 &  0.576 & 2.2$-$3.2 & 24 & 1.3 & 1.5\\ 
 LP 226-48 & 6743 & 0.485 & N/A & 2.1 & 38 & 1.3 & 1.0\\
 G183-35 & 6880 & 0.794 & 3.98 & 8.6$-$10.4 & 66 & 3.2 & - \\
 \hline
\end{tabular}
\end{center}
\end{table*}

Based on these atmospheric parameters, we construct a grid of magnetic spectra using an offset dipole model to determine the viewing angle $i$, magnetic field strength $B_{d}$, and dipole offset $a_{z}$. Figure 1 of \citet{Achilleos89} shows the 3D model for this geometry, and Figure 4 of \citet{Bergeron92} demonstrates the magnetic model spectra when varying one of these parameters and holding the other two constant. As the field strength increases, the Zeeman-split components increase in separation from the central H$\alpha$ line. Changing the dipole offset $a_z$ tends to affect the depth of the split line profiles. \citet{Bergeron92} demonstrated that it is not possible to constrain the viewing angle from the line profiles alone. 
As a result, we only allow $B_{d}$ and $a_{z}$ to vary as free parameters and manually change inclination to search for the best fit model. Our grid includes various inclination values, but we set the inclination to a fixed value before we begin the fitting process and adjust it if no acceptable solution is found using that inclination value. 

\citet{Rolland15} found that the only way to match the line depths in their spectra was to assume each object was in an unresolved binary with a DC WD, effectively diminishing the depths of the line profiles. Though we assume most of  our targets are isolated, we do include this DC offset in our analysis to match line depths between the spectra and our model fits. It is impossible to match the observed line profiles without the addition of this DC component. Our lowest mass targets (LHS 1243, G160-51, and LP 226-48) are likely in binary systems given that they are too low mass ($M<0.5~M_\odot$) to have formed via single-star evolution. However, this does not rule out the possibility of an inhomogeneous atmospheric composition for these low mass objects. The line profiles simply cannot be fit by magnetic pure H atmosphere models, which is true for all targets in our sample. The new flux when including the DC offset is given by Equation \ref{eq1}:
\begin{equation}
{\rm    Total~Flux = Flux_{DA} + DC~offset * Flux_{(DA,\lambda)}}
    \label{eq1}
\end{equation}

Here, ${\rm Flux_{DA}}$ is the initial flux of the white dwarf. The DC offset term is given in Table \ref{tab2} and is multiplied by ${\rm Flux_{DA}}$ at a given wavelength, $\lambda$, which is then added back to the original flux to obtain the new flux, Total Flux. Note that for G183-35, \citet{Kilic19} allowed the effective temperature to vary in their fits in order to match the depths of the absorption features. Hence there is no corresponding DC offset given for G183-35.

We list several relevant parameters for each white dwarf in Table \ref{tab2}, including the effective temperatures and masses from the photometric fits \citep{Caron23}, as well as the best fit magnetic field strength and rotation period when applicable. We also calculate the transverse velocity and list the cooling ages from the literature, which will be important in the conclusions.

\subsection{LHS 1243}

\begin{figure}
 \centering
 \includegraphics[height=2.2in,clip=true,trim=1.8in 0in 2in 0.5in]{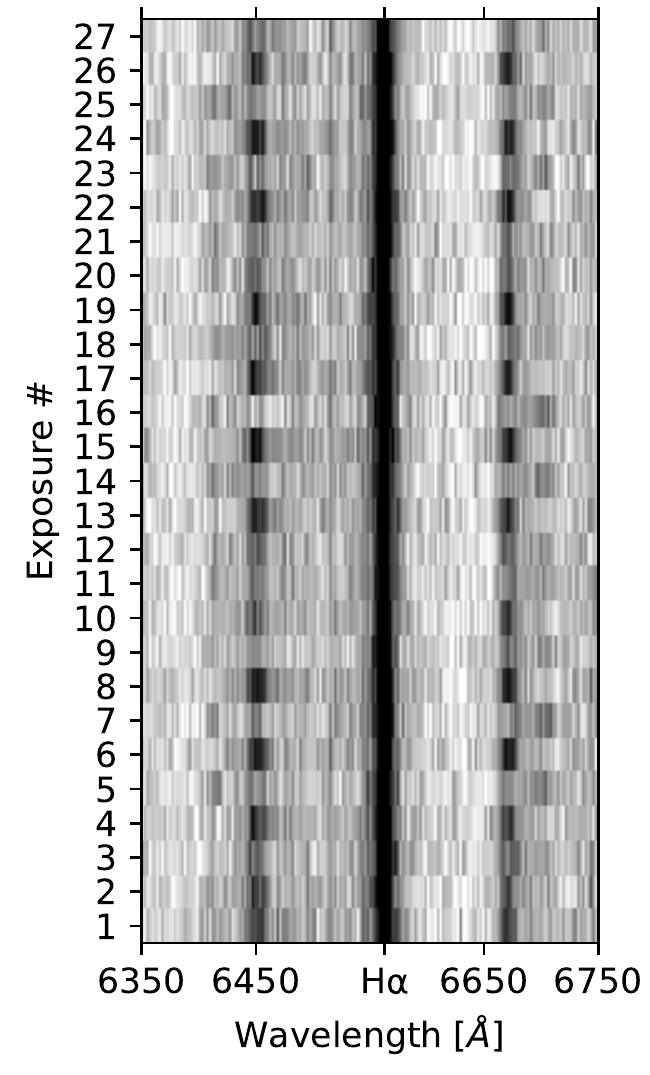}
 \includegraphics[height=2.2in,clip=true, trim=1.2in 0in 1.4in 0.5in]{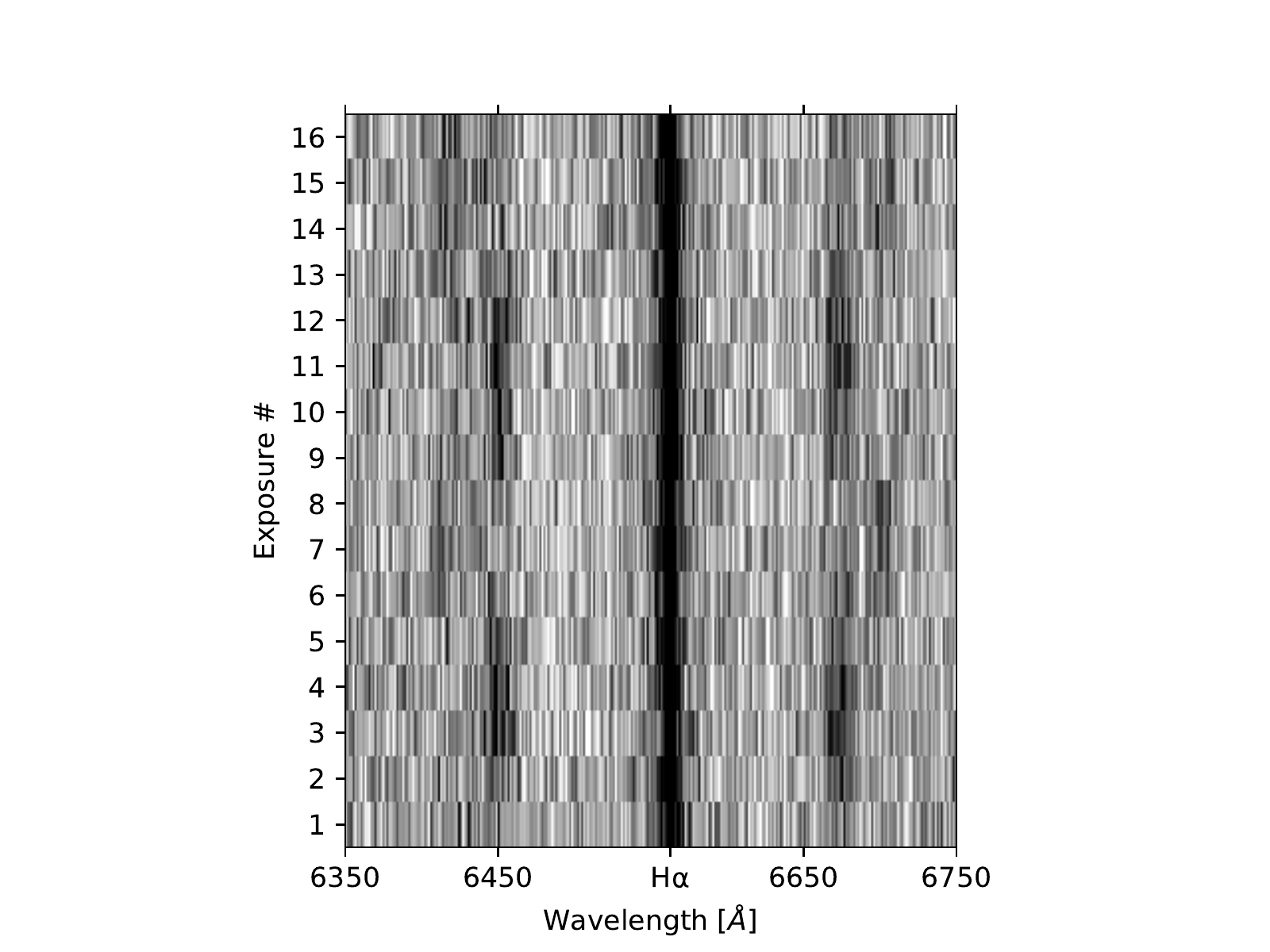}
 \caption{{\it Left:} Gemini time-resolved spectroscopy for LHS 1243 over 2.3 hours. Each exposure is 5 min long. Variations are clear but only hold for one exposure at a time. {\it Right:} Magellan time-resolved spectroscopy over 26 minutes ($16\times80$ s exposures). Here the starting split line position holds for $\sim$4 exposures, then shifts for $\sim$4 exposures, then shifts back.}
 \label{2}
\end{figure}

Figure \ref{2} shows the trailed spectra for LHS 1243. Our Gemini observations, shown in the left panel, revealed variations in the split lines. Here we see three split components, with the outer pair abruptly changing wavelength positions from exposure to exposure. Both lines shift away from the central line by $\sim$30\AA\  before returning to their original location. Given that these shifts are not continuous, we were unable to constrain the period based on the 5 min long exposures from Gemini alone, though it is clear the rotational period should be on the order of $\sim$10 min. Since these exposures were too long to constrain the period, we obtained follow-up spectra at Magellan with 80 s exposures, shown in the right panel of Figure \ref{2}. The lines still seem to shift abruptly but hold their position for multiple exposures compared to our Gemini data. 

\begin{figure}
 \centering
 \includegraphics[width=3.1in,clip=true,trim=1.1in 0.6in 1in 0.6in]{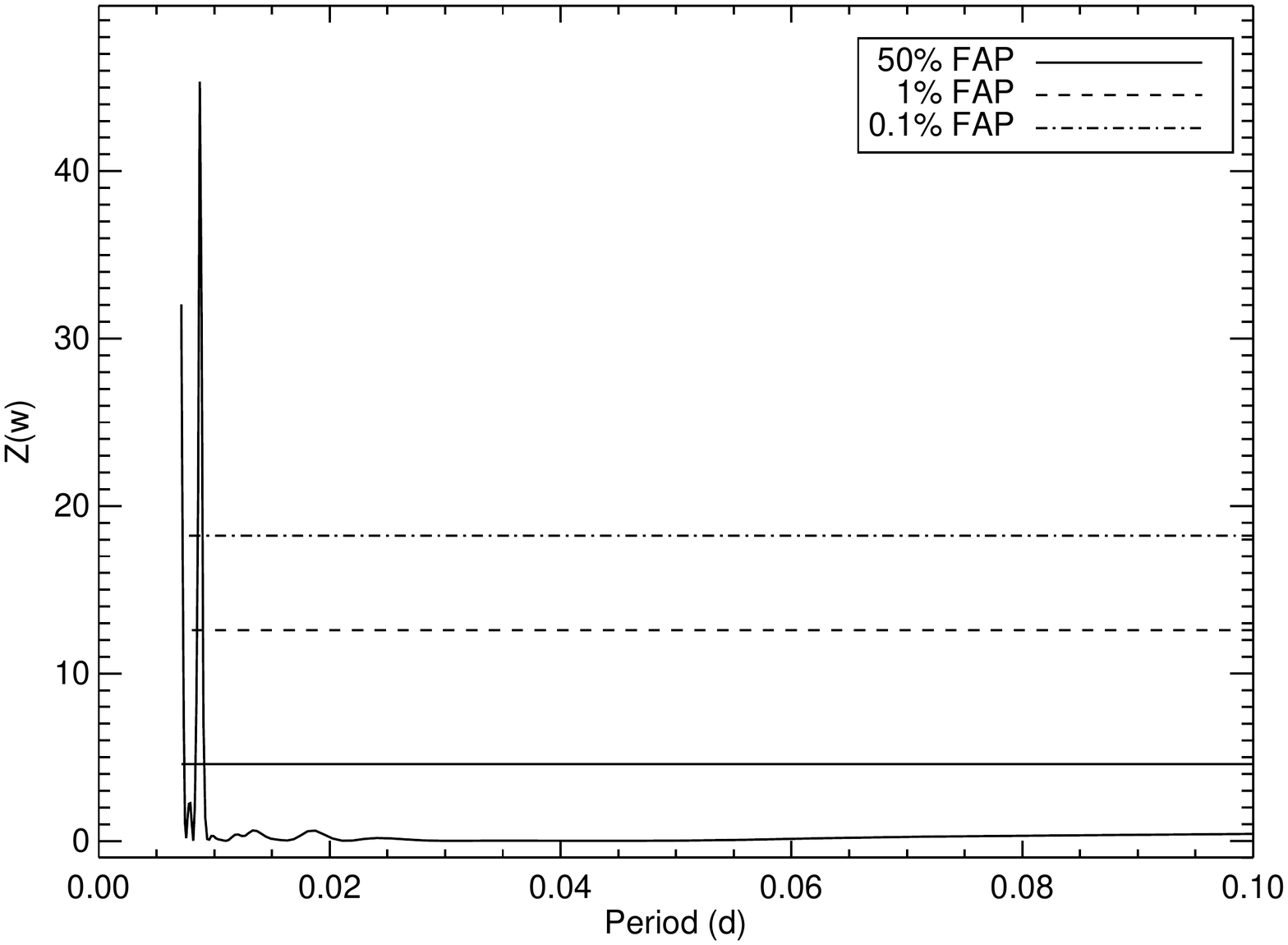}
 \includegraphics[width=3.1in,clip=true,trim=1.1in 0.6in 1in 1in]{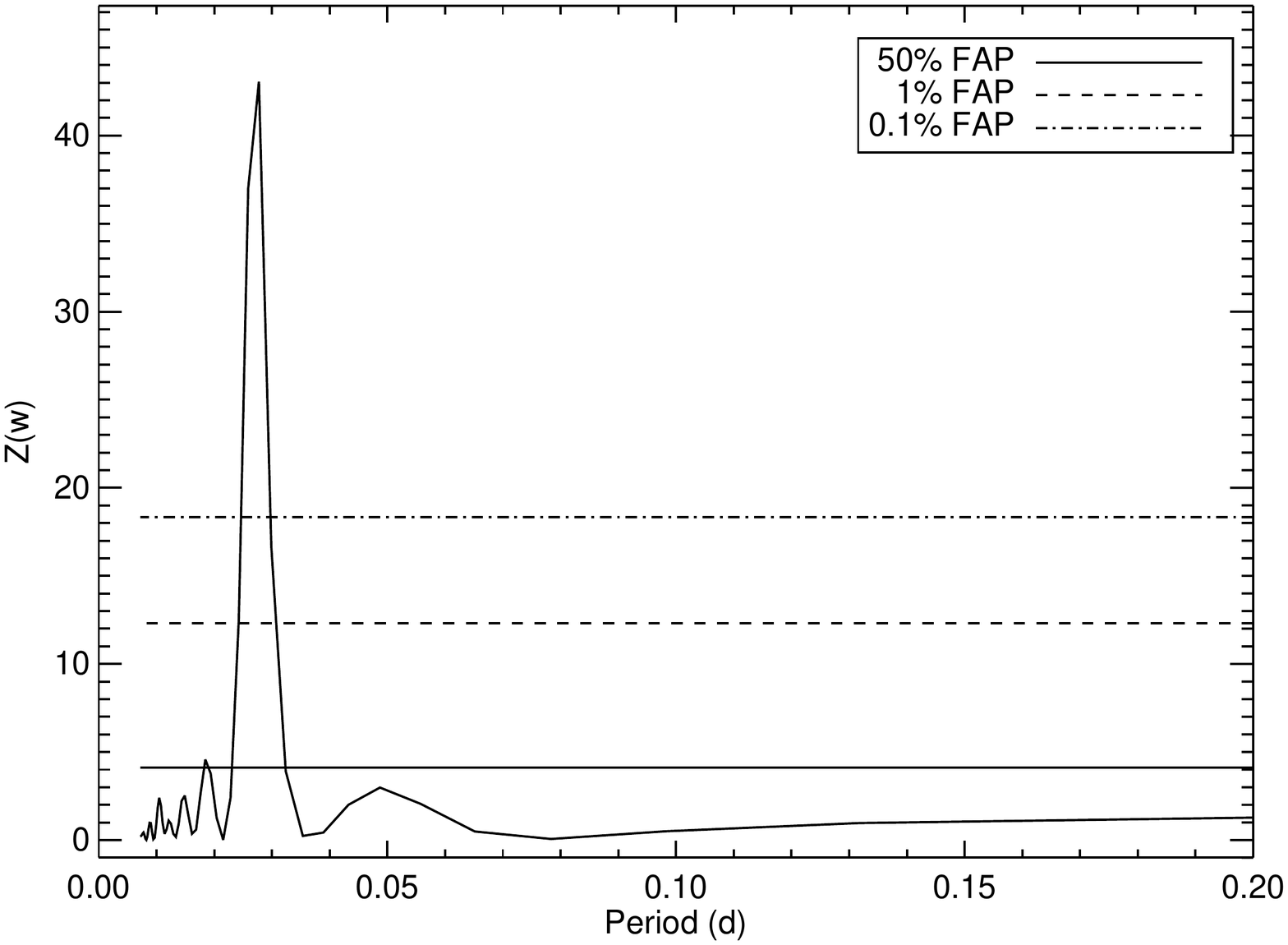}
 \includegraphics[width=3.1in,clip=true,trim=1.1in 0.6in 1in 1in]{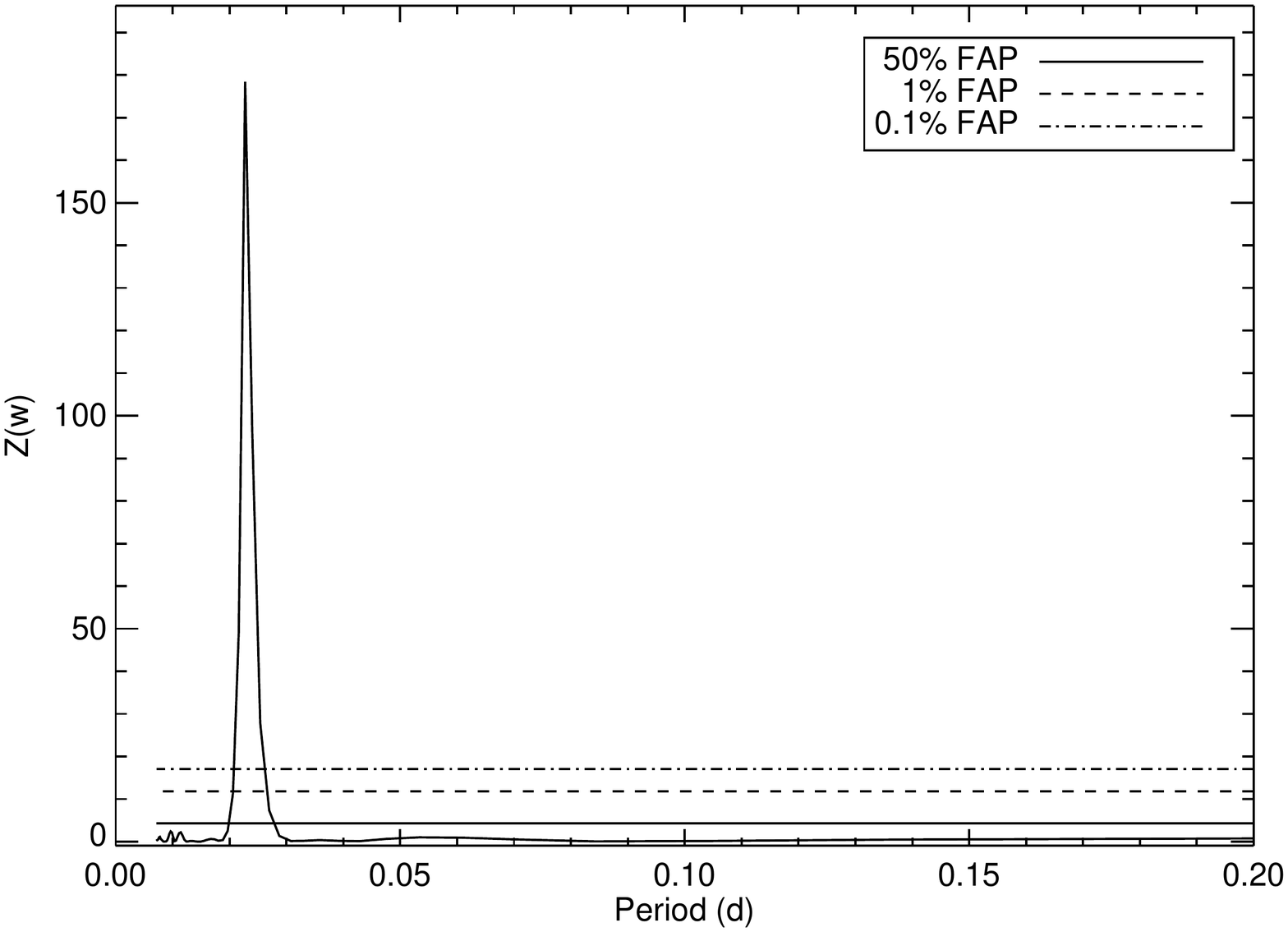}
 \caption{Lomb-Scargle periodograms for our three targets that we could constrain the rotational period for. In all three cases, the period is well above the 0.1\% False-Alarm Probability limit. From top to bottom: LHS 1243, LHS 2273, and PM J15164+2803.}
 \vspace{-0.5cm}
 \label{3}
\end{figure}

In Figure \ref{3} (top left) we confirmed the periodicity by using the line centers from the central H$\alpha$ component to generate a Lomb-Scargle periodogram with the Magellan data. We used the line centers in the first spectrum as the "rest wavelength" and obtained a short rotational period of 12.96 $\pm$ 0.2 minutes, with a low velocity semi-amplitude of $9.7 \pm 1.6$ km/s. This gives a mass function of $f=(8.5 \pm 4.3) \times10^{-7}~M_{\odot}$ and a minimum mass companion of $0.005\pm 0.001~M_{\odot}$. Given that the outer components of the split H$\alpha$ line also vary at the same period, the observed variations in the line center is clearly not due to binarity, but because of a complex magnetic field structure resulting in the line position variations as the white dwarf rotates. Note that LHS 1243 could still be in a binary system since it has a relatively low mass of $M=0.43~M_{\odot}$. 

\begin{figure}
 \centering
 \includegraphics[width=3.2in,clip=true,trim=1.8in 1.35in 2in 1.5in]{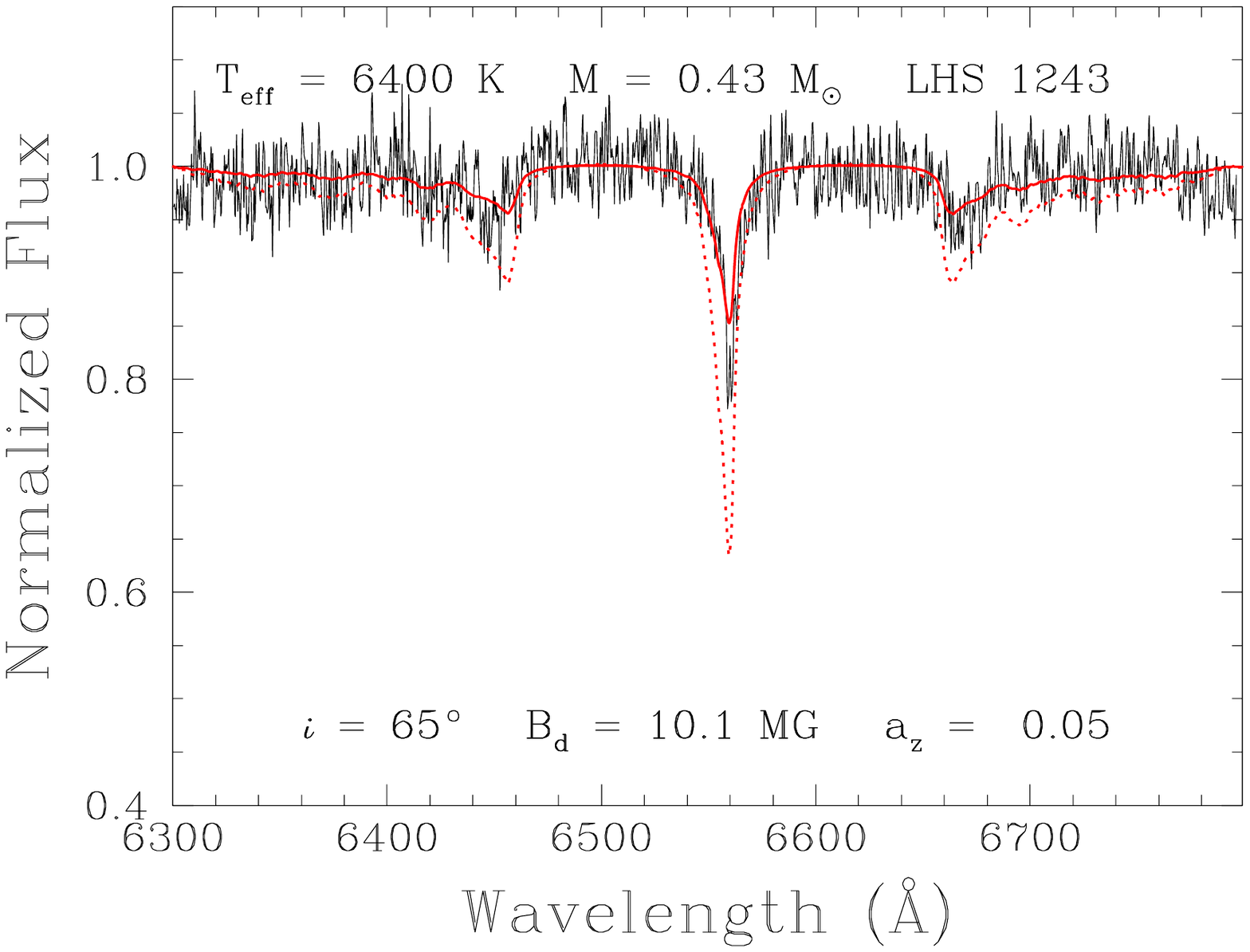}
 \includegraphics[width=3.2in,clip=true,trim=1.8in 1.35in 2in 1.5in]{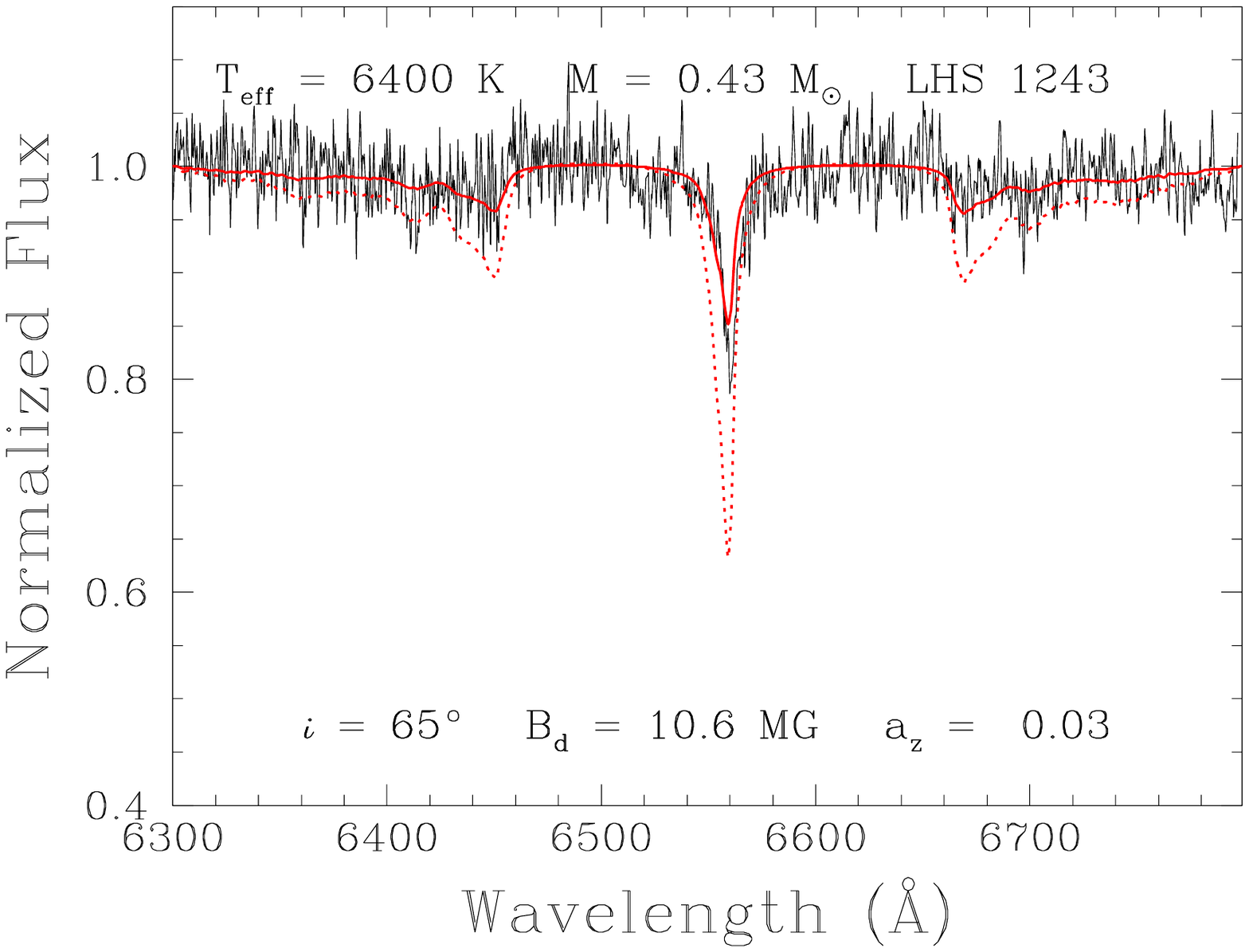}
 \caption{Best fits (solid red line) to two separate spectra of LHS 1243 showing the differences in line position and the resulting effect on the dipole field strength. Note that while the offset value is also different and would change the output model, these values are within the uncertainties whereas the magnetic field strengths are not. The dotted red line shows the same model but without the DC offset included. This results in deeper H$\alpha$ lines that cannot match the observed lines.}
 \label{4}
\end{figure}

Figure \ref{4} shows the model fits to two of our spectra for LHS 1243. Without incorporating the DC offset (shown as the dotted line), all 3 components of the H$\alpha$ line are deeper than observed, which would have a significant impact on the best-fit solution.
This target was previously studied by \citet{Subasavage07} who used Johnson $V$, Kron-Cousins $RI$, and 2MASS $JHK$ magnitudes to determine its $T_{\rm eff}$ and $\log{g}$. Their composite spectrum for this object found a solution of $i = 65^{\circ}$, $B_{d} = 9.5 $MG, and $a_{z} = 0.06$. Given the degeneracies in fitting composite spectra of rapidly rotating magnetic white dwarfs, we assumed the same viewing angle and a positive dipole offset in our fits. $B_{d}$ is slightly larger in our model fits, but given the rapid rotation in this system, our constraints on the $B-$field strength are superior.

\begin{figure}
    \centering
    \includegraphics[width=3.7in,clip=true]{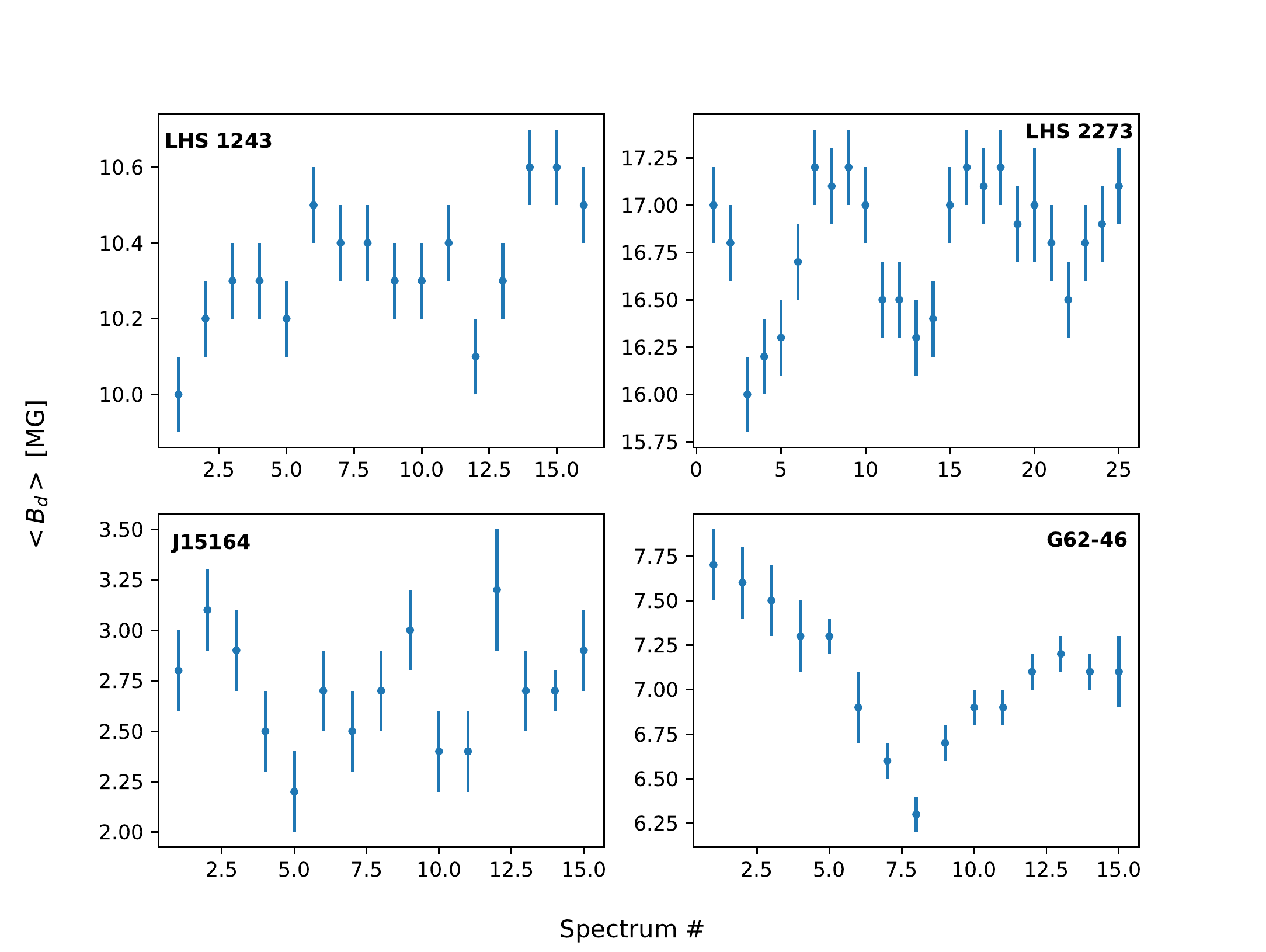}
    \caption{The best fit $B_d$ of each spectrum for our four targets that show spectroscopic variability. The changing position of the Zeeman-split H$\alpha$ lines in the spectra results in a different $B-$field strength. From top left to bottom right: LHS 1243, LHS 2273, PM J15164+2803, and G62-46.}
    \label{5}
\end{figure}

Figure \ref{5} (top left) shows how the best fit $B_{d}$ varies across the Magellan observations. For the first five exposures, the split components are closer to the central line in the trailed spectra, resulting in a lower $B_{d}$ solution ranging from 10.0 to 10.3 MG. At the sixth exposure, the split lines move further from the central line, resulting in a larger $B_{d}$ solution of 10.5 MG. The field strength remains high for several exposures before dropping back down to around 10.3 MG for multiple exposures, and then increasing to 10.6 MG for the remaining spectra.

\subsection{LHS 2273}

\begin{figure}
 \centering
 \includegraphics[width=2.6in,clip=true,trim=6.5in 0.4in 6.5in 0.5in]{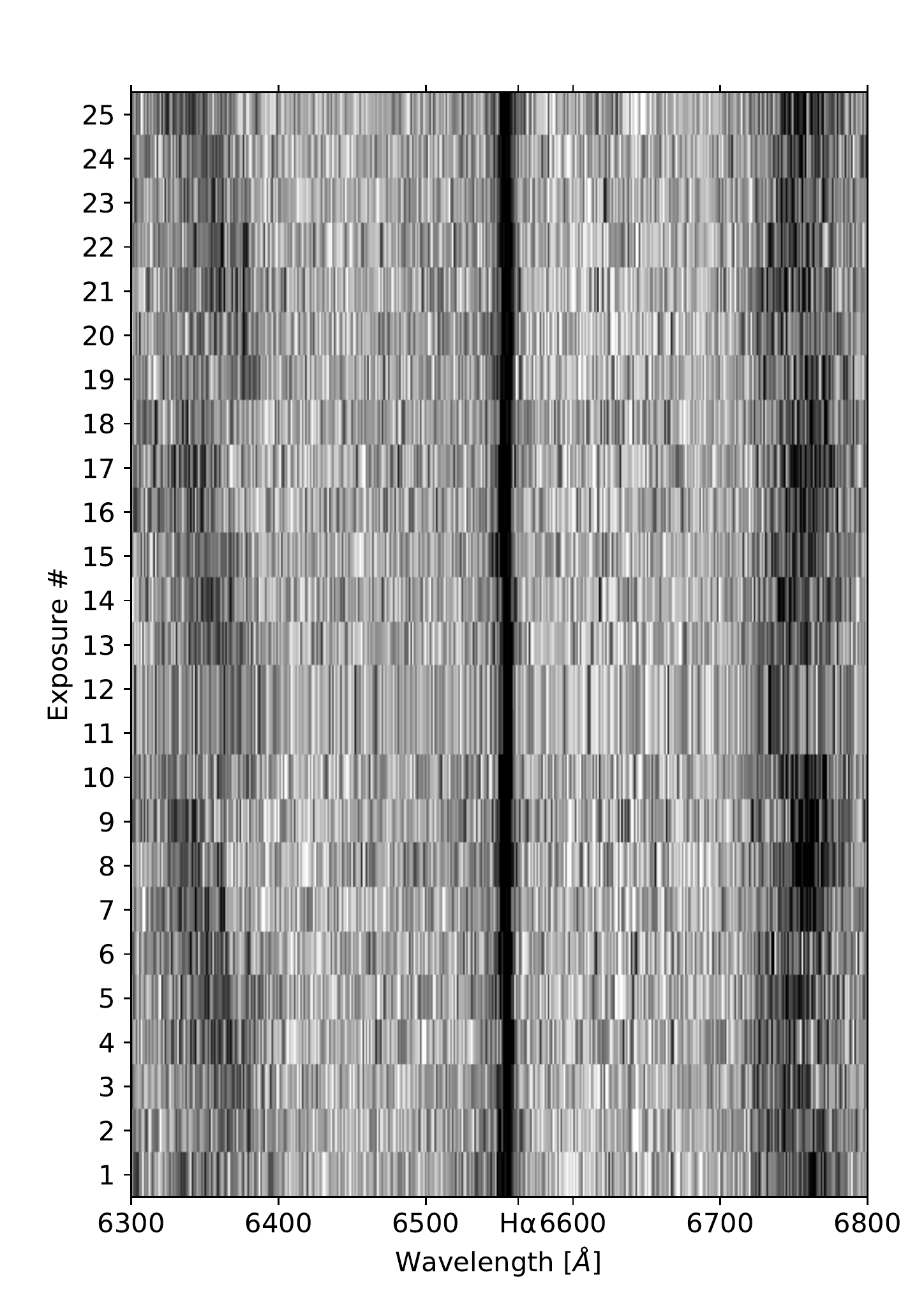}
 \vspace{-0.2cm}
 \caption{Time-resolved Gemini spectra for LHS 2273 over 1.85 hours. The Zeeman split components vary over a period of 39 min.}
 \label{6}
\end{figure}

Figure \ref{6} shows the trailed spectra for LHS 2273. Here the variations in the line positions are more gradual compared to LHS 1243. The total shift is $\sim$60\AA\  but occurs over the span of nine exposures, where each exposure is 4.13 min long. Figure \ref{3} (top right) shows the Lomb-Scargle diagram for LHS 2273 based on the shifts in the central H$\alpha$ line. Based on this and also the variations seen in the split components, the best-fit period is $38.88 \pm 2.53$ min. 

\begin{figure}
 \centering
 \includegraphics[width=3.2in,clip=true,trim=1.8in 1.35in 2in 1.5in]{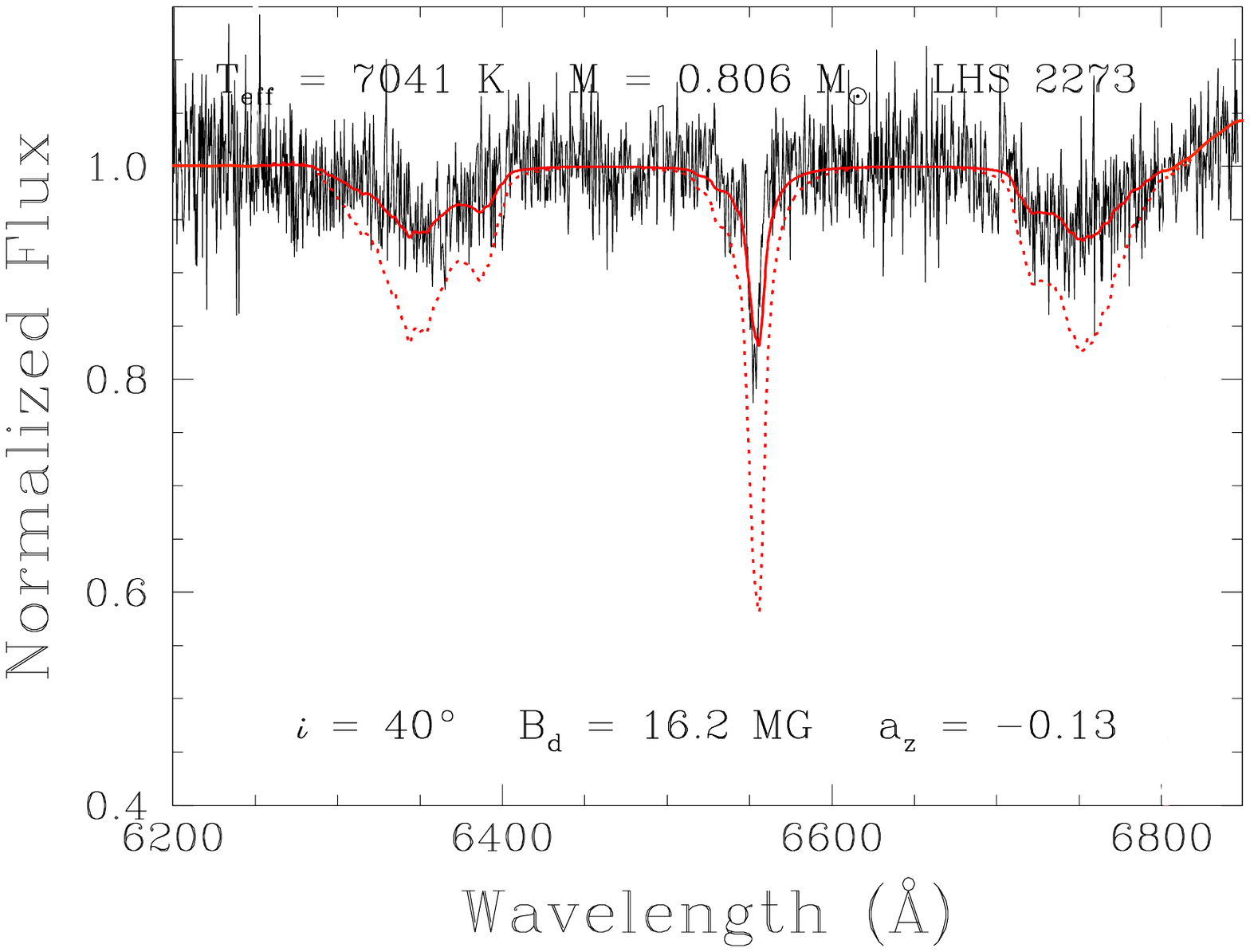}
 \includegraphics[width=3.2in,clip=true,trim=1.8in 1.35in 2in 1.5in]{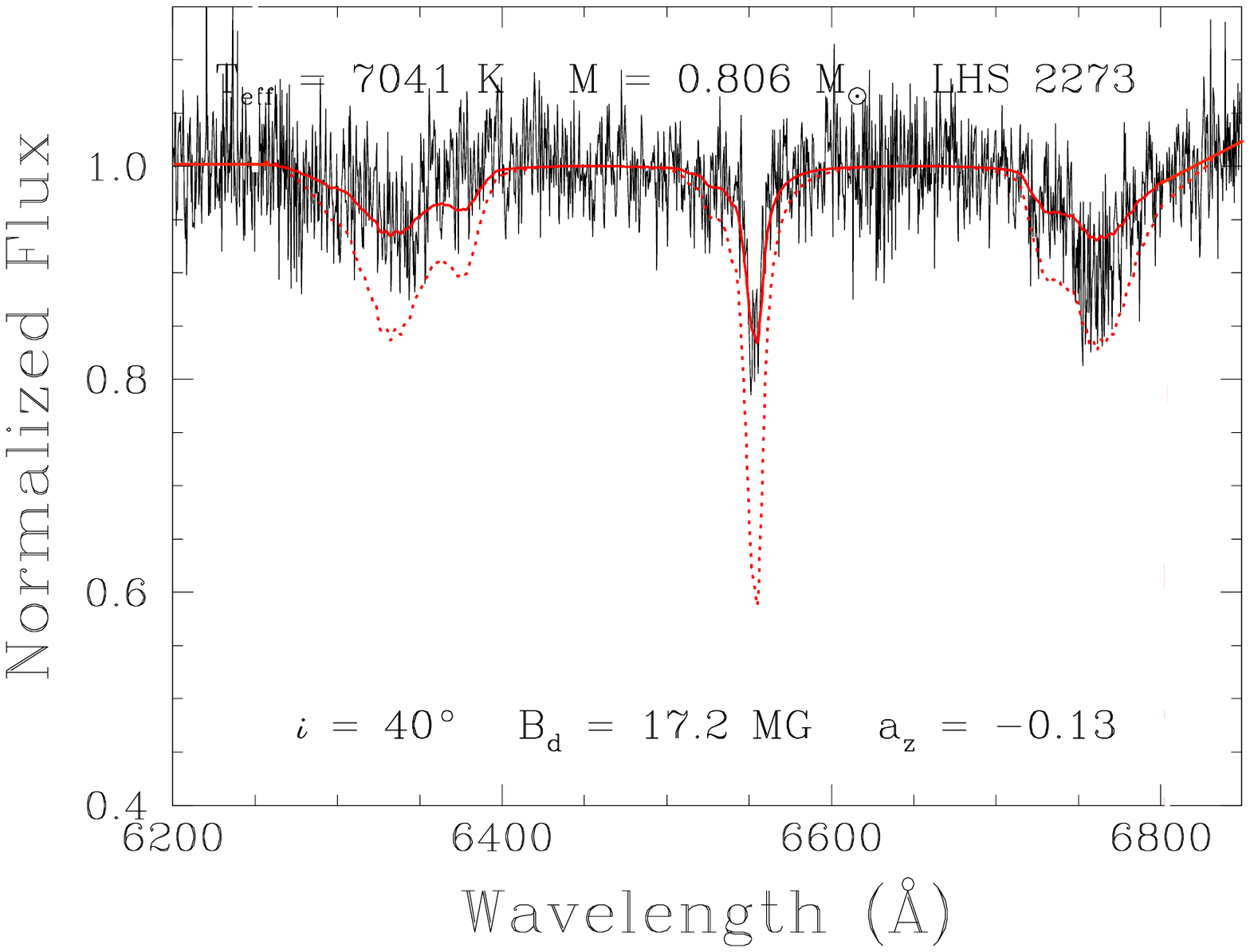}
 \caption{Representative fits to two of our LHS 2273 spectra.}
 \label{7}
\end{figure}

Figure \ref{7} shows fits to two of the spectra for LHS 2273. Here the $B-$field strength changes slightly from 16.0 to 17.2 MG.
Figure \ref{5} (top right) displays how the best fit magnetic field strength varies over our observations, overall matching the pattern seen in the trailed spectra as the components oscillated from shorter to longer wavelengths. The split lines start out far from the central line, resulting in a higher $B_{d}$ solution of $\sim$17 MG. The lines then quickly move closer to the central line, dropping the solution to 16 MG. Over the next several spectra, the split lines gradually move away from the central line, resulting in a gradually increasing $B_{d}$ solution, reaching 17.2 MG. We were able to capture multiple rotations with our Gemini data, so the $B_{d}$ strength continues to gradually oscillate between 16.0 and 17.2 MG and allows us to easily constrain the period.

\citet{Bergeron97} previously fit this target using a centered dipole model and found $B_d=18$ MG and $i=60^{\circ}$. However, the predicted Zeeman components were much deeper than observed, hence the initial prediction of an unresolved DC companion. Given that we are using an offset dipole model, and we include the unresolved DC companion to better fit the fluxes of the Zeeman components, our best-fitting model parameters differ slightly.   
\cite{Hardy23} used the SDSS spectrum of this target to fit the same three parameters and found $B_d=19.18 MG$, $a_z=0.30$, and $i=46^{\circ}$. The slight difference in the magnetic field strength and the notable discrepancy in $a_z$ are not surprising given the rapid rotation of this object, the degeneracies in fitting magnetic WD spectra, and that the SDSS spectra are composite of multiple 15 min long exposures.

\subsection{PM J15164+2803}

\begin{figure}
 \centering
 \includegraphics[width=3in, clip=true, trim=2.0in 0in 2.0in 0.5in]{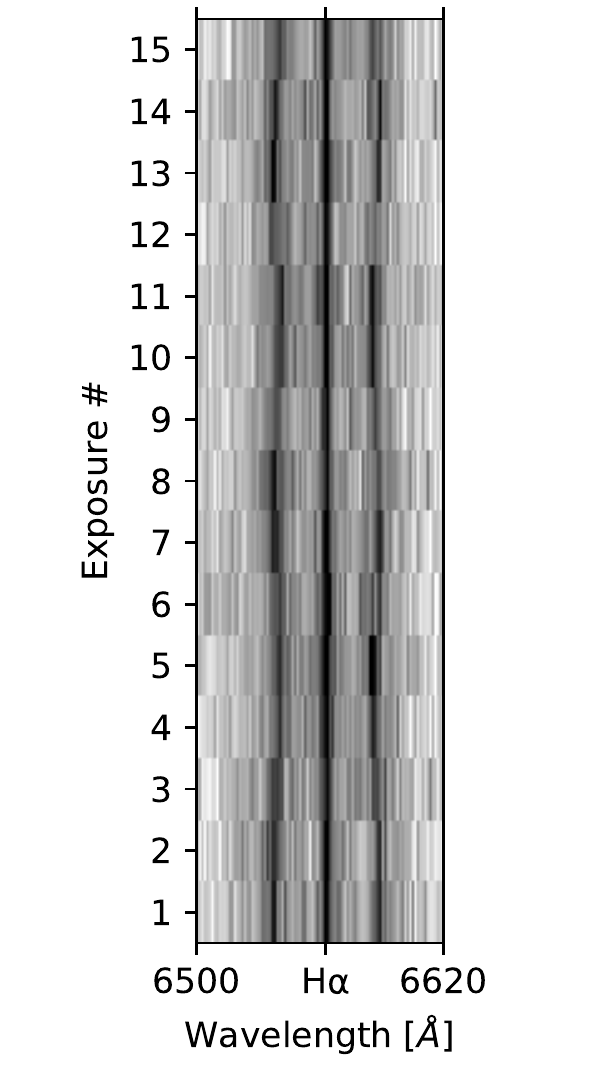}
 \caption{Gemini time-resolved spectroscopy for PM J15164+2803. We only plot the longer exposure times here given the higher signal-to-noise. These observations were taken over 1.33 hours.}
 \label{8}
\end{figure}

Figure \ref{8} shows the trailed spectrum for our final target we could constrain the period for, PM J15164+2803. The oscillations are clear but the magnetic field strength is low, resulting in smaller overall wavelength shifts of $\sim$10\AA. However the depths of the H$\alpha$ lines change rapidly as evident by the contrast in darker and lighter pixels from spectrum to spectrum. This does not have a large effect on the $B_d$ solution but does change $a_z$ which should be constant throughout the spectra. 

\begin{figure}
 \centering
 \includegraphics[width=3.2in,clip=true,trim=1.8in 1.35in 2in 1.5in]{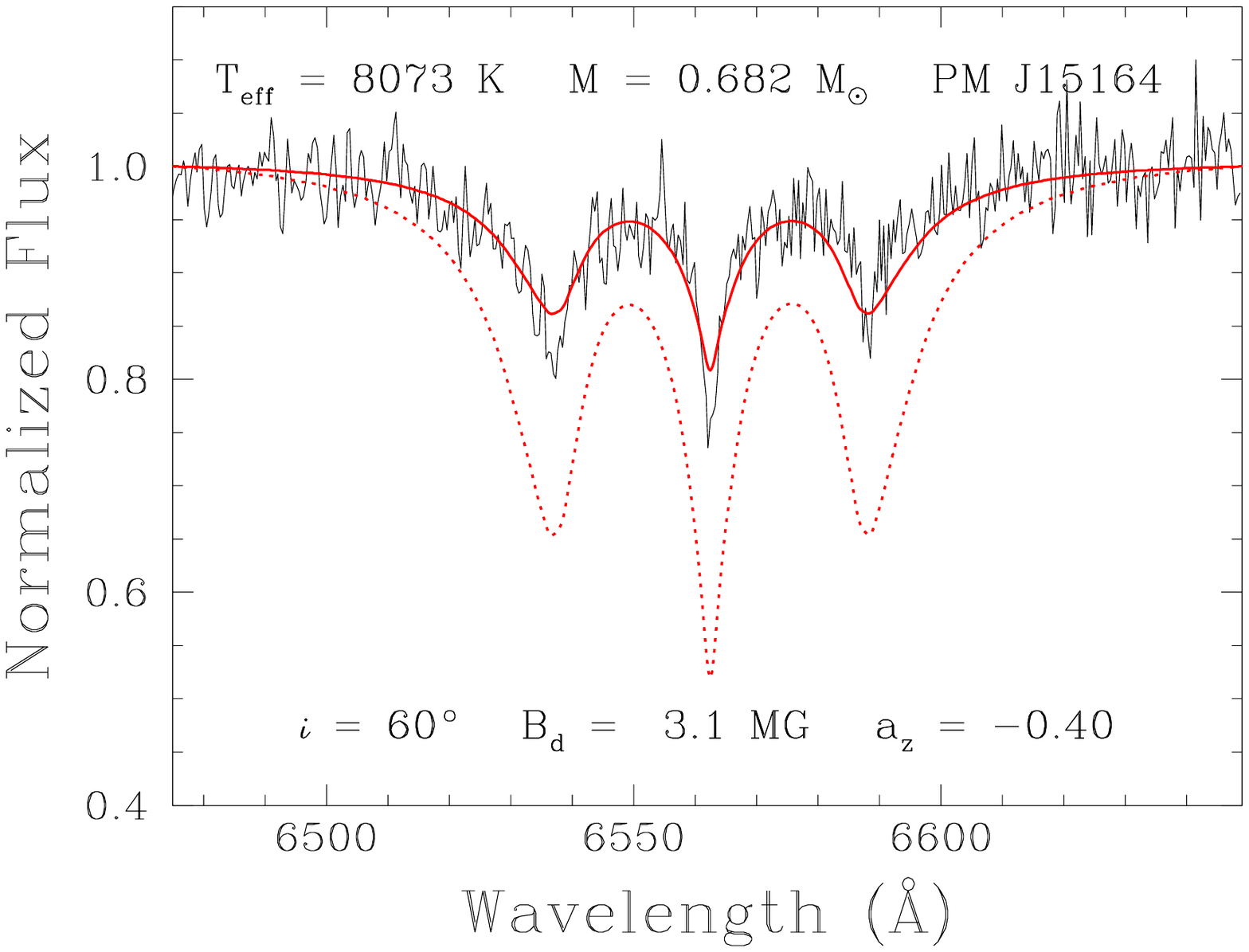}
 \includegraphics[width=3.2in,clip=true,trim=1.8in 1.35in 2in 1.5in]{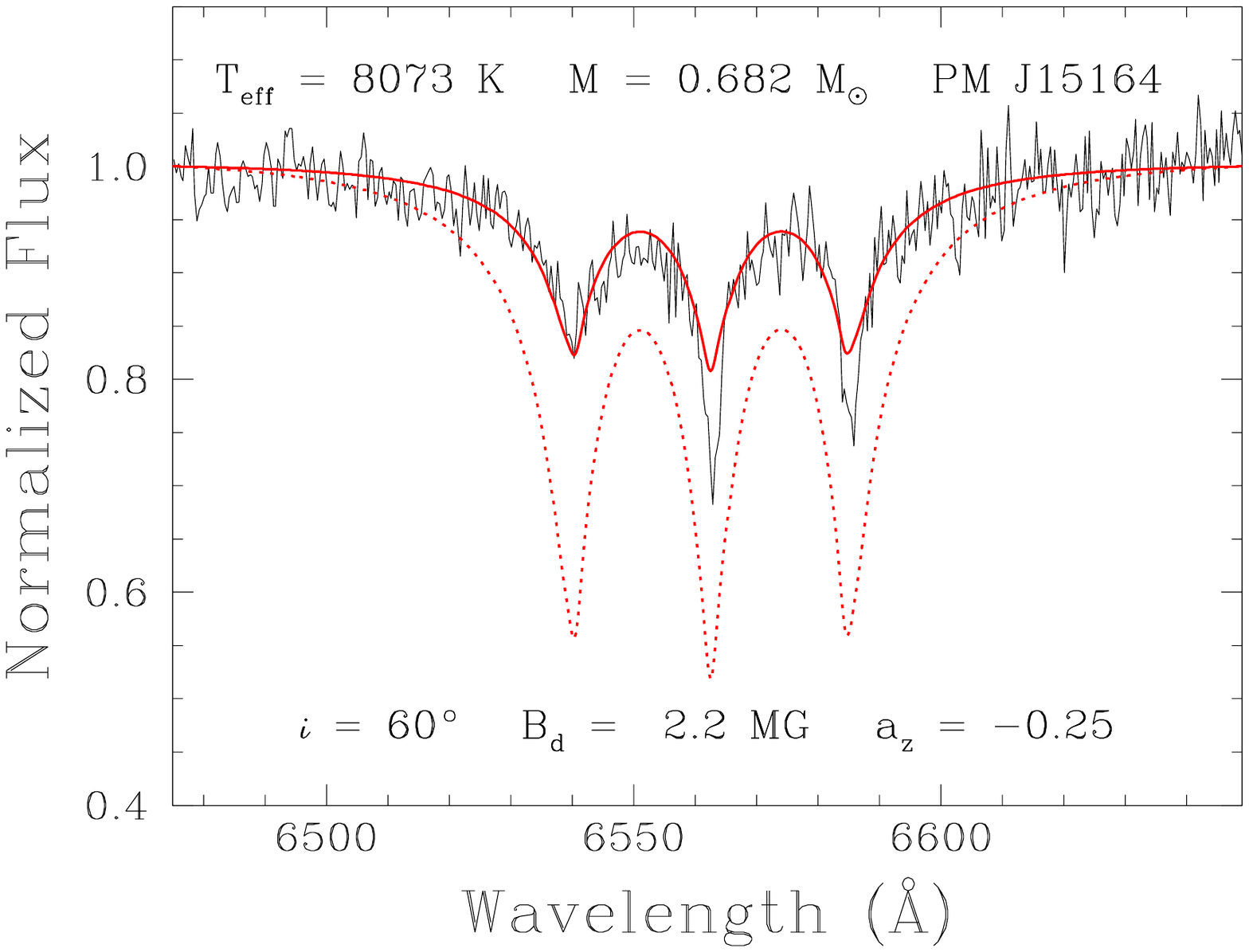}
 \caption{Best fits to two of our PM J15164+2803 spectra. Here the depths of the H$\alpha$ lines vary significantly, resulting in different $a_z$ solutions.}
 \label{9}
\end{figure}

Figure \ref{3} (bottom) shows the Lomb-Scargle diagram for PM J15164+2803 based on the split components of the H$\alpha$ line. The best-fit
period is $34.56 \pm 1.87$ minutes. Figure \ref{9} shows fits to two of the spectra for this target. The most notable feature in
this figure is that our models have trouble matching the depth of the central H$\alpha$ component.

Figure \ref{5} (bottom left) shows the magnetic field variations, again matching the oscillations in the trailed spectra for this object. The wavelength changes in the trailed spectra are relatively small so the $B_d$ solution similarly changes slightly, though the variations are clear, ranging from 2.2 to 3.2 MG. Here the split components start further from the central line, so the $B_d$ solution is larger, $\approx2.8$ MG. The lines then begin to shift closer to the central line, decreasing the fit $B_d$ to $\approx2.2$ MG. Again we were able to observe multiple rotations of this target, allowing us to easily constrain the period and observe the changes in the best fit $B_d$. 

\cite{Hardy23} also analyzed this target, but they found that the absorption lines were too shallow to fit their models and that no combination of parameters could reproduce the observed spectrum. Though they did not definitively arrive at a specific solution, they suggested rapid rotation of an inhomogeneous atmosphere as the cause. Given that the SDSS spectra are combinations of several 15 min long exposures and that the rotation period of this object is about 35 min, it is not surprising that \cite{Hardy23} could not fit the SDSS spectrum.

\subsection{G62-46}
\begin{figure}
 \centering
 \includegraphics[width=3.2in,clip=true,trim=1.1in 0in 1.1in 0in]{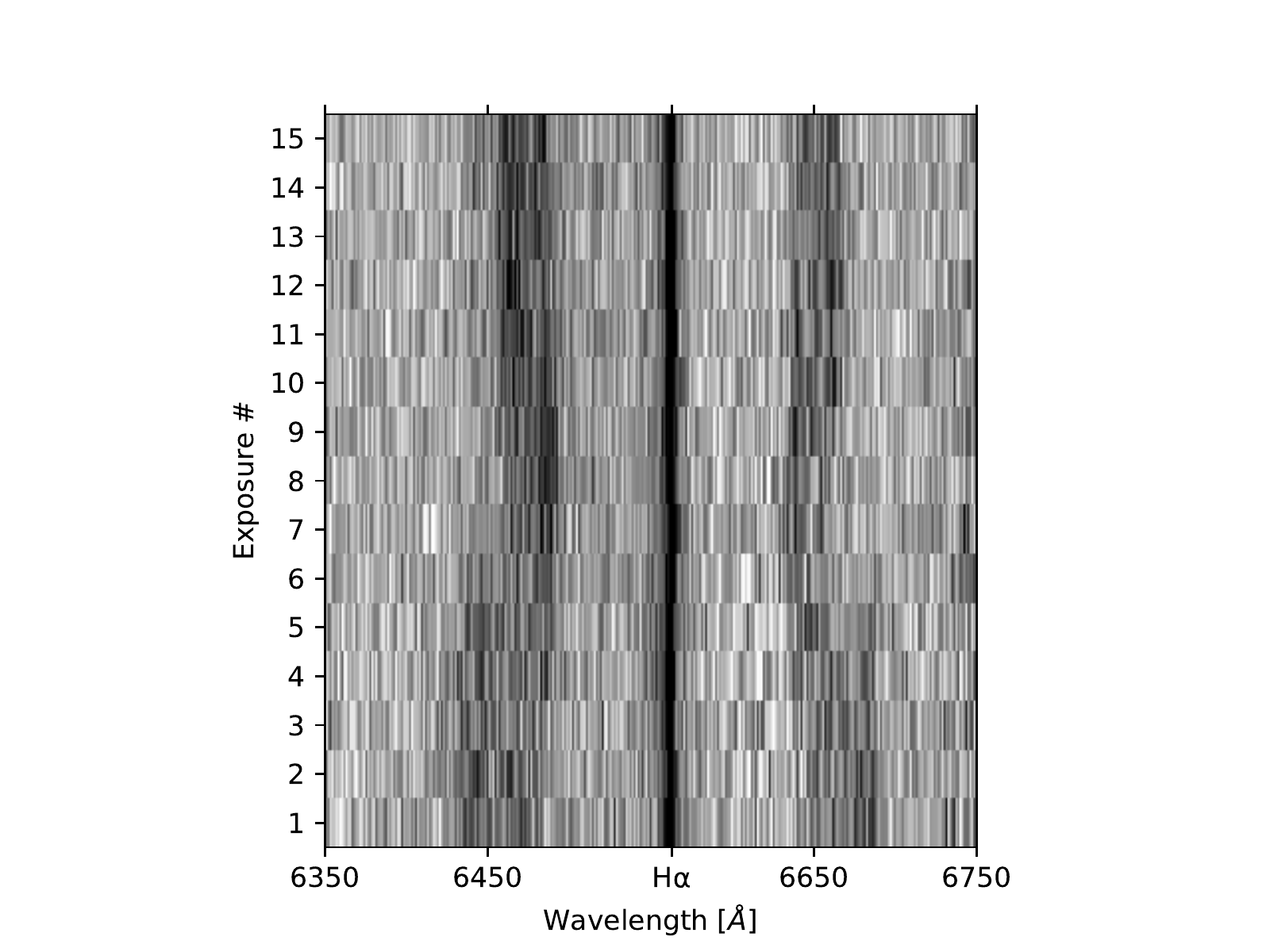}
 \caption{Trailed spectra for G62-46. The changes in position are clear, but our observation span of 1.33 hours was too short to observe a full rotation. Regardless, this target is likely to be a fast rotator.}
 \label{10}
\end{figure}

Figure \ref{10} shows the trailed spectra for G62-46. Here we were only able to capture $\sim$1 full rotation so we cannot confidently constrain the period. However, the variations in the split lines are clear, so this target is still likely to have a rotational period on $\sim$hour timescales. Here we see wavelength shifts of $\sim$30\AA.

\begin{figure}
 \centering
 \includegraphics[width=3.2in,clip=true,trim=1.8in 1.35in 2in 1.5in]{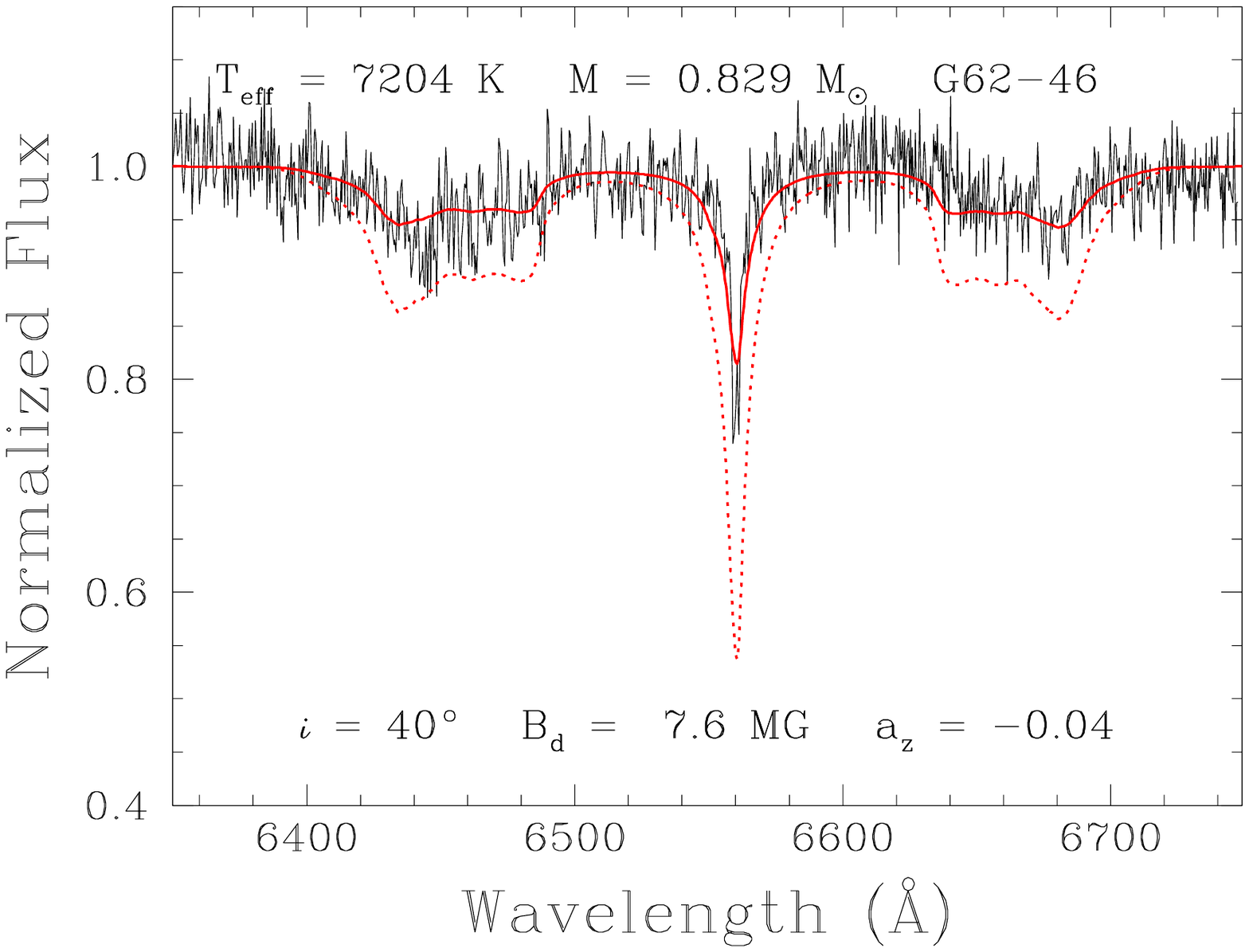}
 \includegraphics[width=3.2in,clip=true,trim=1.8in 1.35in 2in 1.5in]{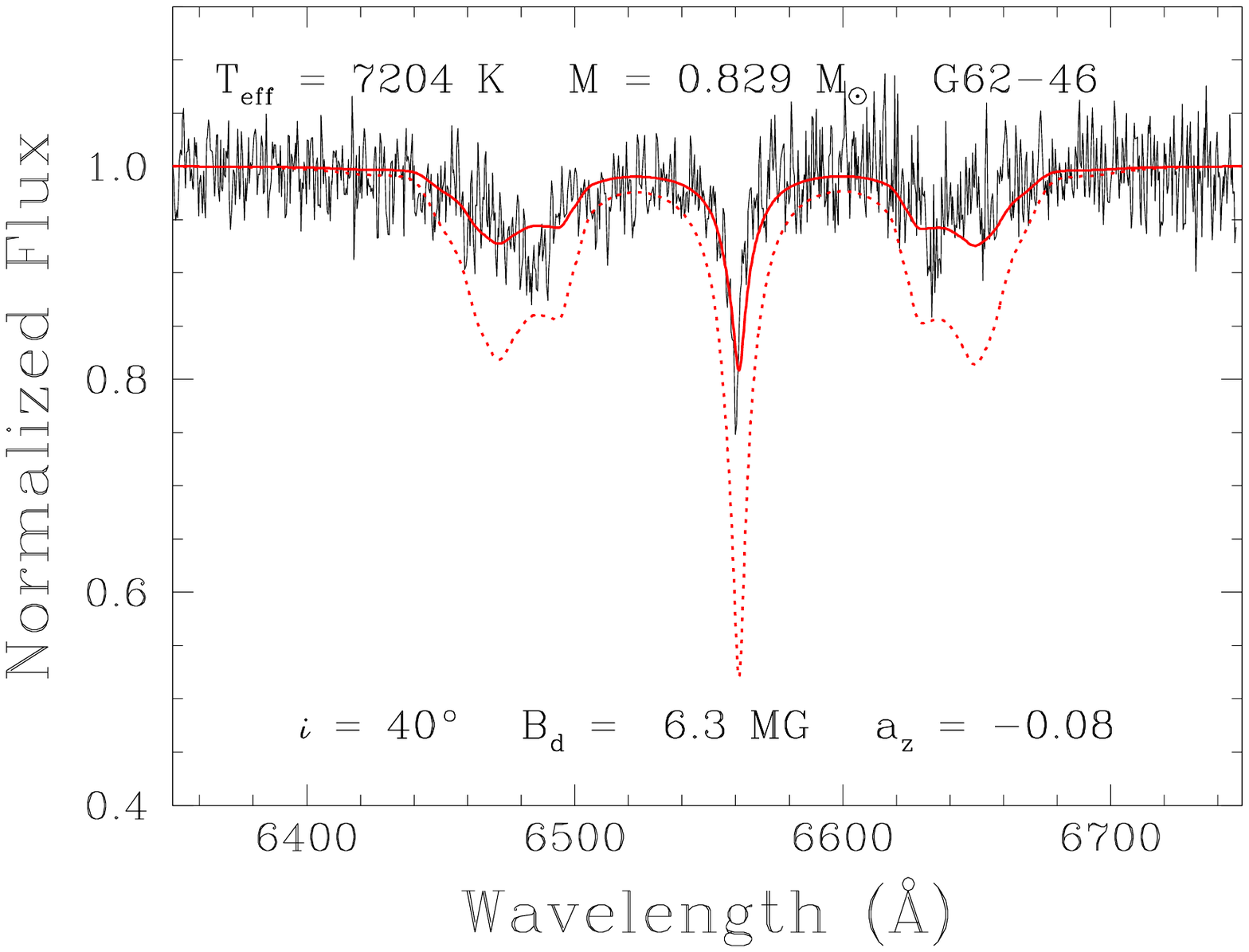}
 \caption{Two example fits for G62-46.}
 \label{11}
\end{figure}

Figure \ref{11} shows two of our example fits for G62-46. While we cannot constrain the rotational period, we still fit each individual spectrum. Figure \ref{5} (bottom right) shows each best $B_d$ solution for our exposures. Once again the patterns seen in the positions of the split components match what we see in the $B_d$ strength. The line positions start further from the central line then gradually reach their closest position halfway through our observations, then move back towards the end. The $B_d$ similarly starts at its largest value of 7.7 MG and reaches a minimum of 6.3 MG halfway through our observations, then increases back.

\subsection{The Rest}

In Figure \ref{12} in the Appendix we show the trailed spectra for our remaining three targets. These three did not show variations in their spectra, so we simply combined the spectra into an average to increase the signal-to-noise, and then fit the combined spectrum. These fits are shown in Figure \ref{13}. The field strengths for these three targets range from 2.1 to 3.2 MG. 

\cite{Hardy23} analyzed LP 226-48 and found $B_d=2.4 MG$, $a_z=0.30$ for $i=50^{\circ}$, whereas we found $B_d=2.1 MG$, $a_z=-0.13$ for $i=30^{\circ}$. The $B-$field measurements are consistent within the errors, but the difference in $a_z$ is likely due to the noticeably different inclination adopted in that work. 

We are unable to produce an adequate fit for G160-51. This target is very different from the rest of the sample; the Zeeman-split components separate from the central H$\alpha$ line, which is deeper and broader on average than the other targets. The low surface gravity obtained from the photometric fit suggests this target is a low-mass 
WD with $M=0.385~M_{\odot}$ and it is likely in a binary system with another DA companion that could explain the unusual H$\alpha$ line profile.

\section{Discussion}

\subsection{Rotational Variability}

Of the 10 WDs in proposed unresolved binaries from \citet{Rolland15}, we now have obtained time-resolved spectroscopy on 8 of them (including G183-35 from \citealt{Kilic19}), 5 of which show variations in their spectral features on the scale of hours or shorter. In this work we successfully constrained the rotation periods for 3 targets by observing the changing position in their Zeeman-split H$\alpha$ lines: LHS 1243, LHS 2273, and PM J15164+2803. Each target has a rotation period less than an hour: 12.96 $\pm$ 0.2, 38.88 $\pm$ 2.53, and 34.56 $\pm$ 1.87 min, respectively. 

We used an offset dipole model to fit each spectrum to see how the average surface magnetic field strength changes as the WD rotates. The strength for each target varies from 10$-$10.6, 16.0$-$17.2, and 2.2$-$3.2 MG respectively. The magnetic field itself does not change strength, rather with the magnetic axis offset from the rotation axis, we see a different magnetic field distribution across the stellar surface as the objects rotate. G62-46 shows clear variations but the rotation period is longer than our observations, though not by much. The remaining three targets did not show any variations in their spectra, so we simply fit one model to the combined spectrum. It is likely these targets have rotation periods much longer than a few hours.  

\subsection{Patchy Atmospheres}

All of our targets require dilution by a hypothetical DC companion (or a likely DA companion in the case of G160-51) in order to match the line profile depths between our models and observations. Without it, our model spectra always produce deeper line profiles that result in poor fits to the observations. However,
there is no evidence of significant radial velocity variations in our data. All of the variations seen in the H$\alpha$ line profiles of our targets can be explained by rotational modulation of a complex magnetic field structure. 

In \cite{Rolland15}, under the assumption of an unresolved binary, all of these objects were found to be within the non-DA gap between 5000 and 6000 K \citep{Bergeron97}. Now with Gaia parallaxes and assuming single objects, our new estimates place them at warmer temperatures between 6400 and 8000 K. It is clear from the discrepancies between the spectroscopic \citep{Rolland15} and photometric temperature estimates \citep{Caron23} and also the discrepancies between the predicted and observed line profiles that these atmospheres are not pure H. In fact, rotation of a magnetic white dwarf with a chemically inhomogeneous surface, similar to the Ap/Bp stars, can explain the spectroscopic variations seen in these stars. Note that the three low mass WDs in our sample are likely in unresolved binaries, but can still have inhomogeneous atmospheres.

GD 323 is an excellent example of a WD with an inhomogenous surface composition, where the strengths of the H and He lines change over
a period of about 3.5 h \citep{Pereira05}. GD 323 is hot enough to show both H and He lines, and therefore is classified a DAB, but
our targets are too cool to show He absorption lines. In the case of GD 323, \citet{Pereira05} argued that an inhomogeneous surface
composition can be the result of the dilution of a thin hydrogen atmosphere with the underlying helium convection zone. 
The presence of magnetic fields on MG scales rules out the same scenario for our targets. \cite{Tremblay15} ran radiative magnetohydrodynamics simulations and found that even a magnetic field of $\sim$50 kG could suppress convection, which was observationally confirmed by \cite{Gentile18}. Given that all of our targets have fields strong than 2 MG, it seems likely that there should be no convective mixing in our targets. Hence, the source of an inhomogenous composition in our targets is currently unclear.

\subsection{Source of Magnetism}

\begin{figure*}
 \centering
 \includegraphics[width=0.7\textwidth,angle=-90]{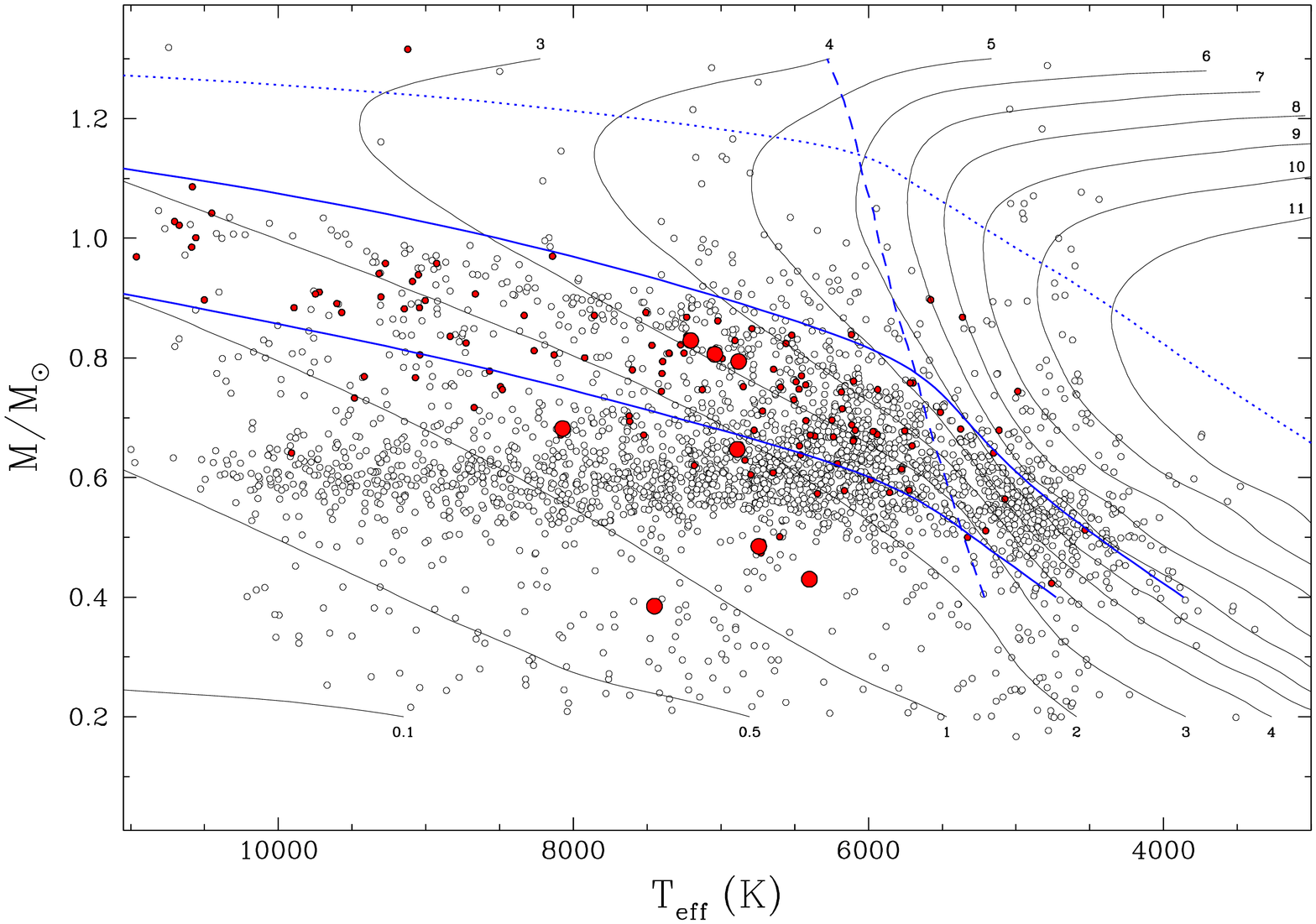}
 \caption{Stellar masses as a function of effective temperature for white dwarfs in the \citet{Caron23} sample. Solid curves are theoretical isochrones, labeled in units of $10^9$ yr, obtained from cooling sequences with C/O-core compositions, $q(\mathrm{He})\equiv M(\mathrm{He})/{M}_{\star }={10}^{-2}$, and $q(\mathrm{H}) = 10^{-4}$. The lower blue solid curve indicates the onset of crystallization at the center of evolving models, while the upper one indicates the locations where 80\% of the total mass has solidified. The dashed curve indicates the onset of convective coupling, while the dotted curve corresponds to the transition between the classical to the quantum regime in the ionic plasma. Our targets are marked in large red circle.}
 \label{14}
\end{figure*}

We first consider the fossil origin of the magnetic field from Ap/Bp stars in our rapid rotators. We use the initial-final mass relation from \cite{Cummings18} to estimate the initial mass of the progenitor stars from our white dwarf masses. LHS 1243 is too low mass (0.430 $M_{\odot}$) to have evolved from single star evolution. It is extremely likely this object underwent some binary interaction in the past, leading to both the short rotation period and the magnetic field. This would involve two common envelope evolutionary phases similar to NLTT 12758 \citep{Kawka17} as well as a dynamo driven by differential rotation \citep{Tout08,Briggs18b}, or the expulsion of the majority of the envelope during a common envelope phase merger. PM J15164+2803 has a mass of 0.68 $M_{\odot}$, which yields an initial mass of 2.41 $M_{\odot}$. Our more massive targets, LHS 2273 and G62-46 (0.806 $M_{\odot}$ and 0.829 $M_{\odot}$), yield similar initial masses of 3.33 and 3.45 $M_{\odot}$. Therefore it is possible for our higher mass targets to have originated from these highly magnetic main-sequence stars. However, \citet{Hermes17} analyzed the rotation rates of pulsating white dwarfs using asteroseismology and found that white dwarfs that evolved from 1.7 - 3.0 $M_{\odot}$ progenitors typically have $\sim$day rotation periods. These progenitors form average mass white dwarfs (0.51 - 0.73 $M_{\odot}$), and all four of our WDs with $M\geq0.68~M_{\odot}$ in our sample (see Table \ref{tab2}) have hour timescale rotation periods. Hence, a fossil origin is unlikely to explain the observed magnetism and rapid rotation in these WDs. 

Crystallization, though still not well understood, could also generate magnetic fields via the generation of a dynamo.
For example, \citet{Ginzburg22} show that fields up to 7$-$10 MG can be generated for a crystallized WD with a spin period of 0.5$-$1 h (see their Figure 4). \citet{Caron23} performed a spectrophotometric analysis of 2880 cool white dwarfs within 100 pc. The vast majority of their magnetic objects lie within or near the region where crystallization occurs in Mass-$T_{\rm eff}$ space, shown in their Figure 17. This suggests a strong correlation between the presence of a magnetic field and core crystallization. 

\citet{Caron23} hypothesized that the transition of most cool, He atmosphere WDs to H dominated atmospheres may be due to the crystallization process. In this scenario, the center crystallizes first and the crystallization front propagates outward, generating a dynamo which creates the magnetic field. This field then inhibits convection, allowing H which was originally diluted in the mixed H/He envelope to diffuse upwards, resulting in a H atmosphere. Hence, this scenario could explain the prevalence of H atmosphere WDs below 5200 K.

We reproduce the stellar mass versus effective temperature diagram for the \citet{Caron23} sample here in Figure \ref{14}, highlighting our sample of magnetic WDs with large red circles. Among our 8 magnetic DA WD targets, only three have begun crystallization. Based on the evolutionary models \citep{Bedard20}, we expect LHS 2273, G62-46, G183-35 to have 17, 18, and 46\% of their cores crystallized. 

Note that the evolutionary models shown in Figure \ref{14} assume equal amounts of C and O in the core. If the oxygen abundance in the core is higher \citep[e.g. $\sim0.8$,][]{Giammichele18,Giammichele22}, crystallization would start earlier because the average charge of the ion and the coupling constant are larger. At $0.7~M_\odot$, a larger oxygen abundance in the core would shift the start of crystallization from $T_{\rm eff}=7000$ K to 8000 K. Hence, the crystallization region (the area between the two blue lines shown in Figure \ref{14}) could encompass almost all magnetic WDs included in this Figure, including all but the three lowest mass WDs in our sample. 

While crystallization could be the driving force for dynamo generation in five of our targets, three of our targets have not begun crystallizing, nor can crystallization explain the rapid-rotation shown in five of our targets. Binary evolution is the most likely candidate for the observed rapid rotation in addition to the source of the magnetic field. We include the discussion of crystallization-driven dynamos given that it could play a role for five of our targets, but we expect that seven of our targets have undergone some form of binary evolution in the past. Additionally, not every WD that lies within the crystallization region in Figure \ref{14} is magnetic, so it would seem crystallization alone cannot produce the dynamo required to generate a magnetic field. It is tantalizing to suggest rapid rotation is also needed to drive up this dynamo and produce magnetism. Only one of our crystallized targets (GD 175) does not show signs of hour timescale rotations, though it could rotate on day-long timescales. Further time-series spectroscopy on these magnetic WDs that lie within the crystallization region could determine what fraction have fast rotation periods and if this rotation aids in producing magnetism in crystallized WDs.

\citet{Bagnulo22} found two distinct MWD populations based on mass: those above 0.75$M_\odot$, which are almost certainly merger products given the high mass and extremely strong magnetic fields ($\sim$ 100 MG); and those below 0.75$M_\odot$, likely the products of single-star evolution and fields that actually grow over time, typically in the first 2-3 Gyr of cooling (see also \citet{Amorim23}). The class of hot ($10000 < T_{\rm eff} < 25000$K) carbon-rich DQs exhibit properties consistent with double-degenerate mergers: large mass \citep[$>0.8~M_{\odot}$,][]{Coutu19}, high transverse velocity, rapid rotation, and magnetism \citep{Dunlap15}. To determine the progenitors of these objects, \citet{Kawka23} conducted population synthesis studies for double degenerate mergers and He star mergers with a WD. They found that the double-degenerate mergers produced white dwarfs with larger delay times, higher mass, and older kinematic ages. These findings are consistent with the hot carbon-rich DQs and strongly support the idea that they form via double-degenerate mergers. 

Binary mergers can explain a significant fraction of ultramassive white dwarfs \citep{Kilic23}, however binary evolution is still relevant even for average mass WDs. Binary population synthesis calculations of \citet{Temmink20} find that the mass distribution of single WDs that form through binary mergers peak at $0.64~M_{\odot}$, with a median at 0.65 and mean at $0.71~M_{\odot}$. \citet{Briggs15} considered only mergers that could generate magnetic WDs and found a mean predicted mass of $0.88M_{\odot}$ for magnetic WDs compared to $0.64M_{\odot}$ for all WDs.  Observational signatures of a single WD that form through binary evolution may include rapid rotation, magnetism \citep{Tout08}, or large tangential velocity. 

Of our eight targets, three of them (LHS 1243, G160-51, and LP 226-48) are too low mass to have formed via single star evolution ($M< 0.5M_{\odot}$) given the age of the Universe. Hence, these three stars must have gone through binary evolution. LHS 1243 also happens to be the fastest rotator in our sample. Of the remaining five targets that are more average in mass, four of them (WD 1026+117, PM J15164+2803, G62-46, G183-35) have magnetic field strengths on MG scales and rotate on minute/hour timescales, strong indicators of past binary interactions. 

We also calculate the transverse velocity of each target using proper motions and parallaxes from Gaia DR3 to see if any of them are outliers in their kinematics compared to the Galactic disk. \citet{Temmink20} showed that the ages of single WDs that form through binary evolution are significantly underestimated, and that individual WDs appear $\sim1$ Gyr younger than they are if binary evolution is ignored. Hence, some of the single WDs that form through binary evolution may appear as outliers in their kinematics. \cite{Kilic20} looked at the velocity distributions for DAs with cooling ages below and above 2 Gyrs. For our five targets that are younger than 2 Gyrs, only one (LHS 1243, the fastest rotator) has a notably large velocity of 53 km/s, faster than $\sim$70\% of other young DAs. All three of our older targets though have larger transverse velocities than $\sim$60\% of the rest of the population older than 2 Gyrs, with LHS 2273 and G183-35 (both fast rotators) in the top $\sim$10 and 20\% respectively. This ultimately leaves one object with no particularly noteworthy qualities, GD 175. This object has a mass of 0.65$M_{\odot}$, an unremarkable transverse velocity, and no signs of rotation. Hence, 7 of the 8 magnetic WDs in our sample show evidence of past binary evolution that would explain their rapid rotation, magnetism, or low mass. 

An emerging class of magnetic white dwarfs dubbed "DAHe" show variable He emission with short rotation periods \citep{Reding23,Manser23}. Their similarities to our sample are noted given the rapid rotation and magnetic field strengths. However, this class of $\sim$25 objects occupy a much narrower range of mass and color, and their formation mechanism is still unclear. Therefore we cannot make any direct connections between that class and our sample besides the idea of binary interactions/mergers. Time-series spectroscopy of additional magnetic white dwarfs across a wider range of effective temperatures and masses will be necessary to detect more systems with variable line profiles and confirm this hypothesis, as well as further advances in the theoretical framework of magnetic field generation.

To summarize, seven of our 8 targets show at least two indicators of binary evolution: magnetism, and either rapid rotation or low mass. LHS 1243 has both a low mass and short rotational period. G160-51 and LP 226-48 do not show signs of rapid rotation but have too low mass to have formed from single-star evolution. LHS 2273, G62-46, PM J15164+2803, and G183-35 are average mass but have rotational periods typical of binary evolution products. Crystallization could play a role in generating a dynamo for five of our targets including GD 175, however this mechanism itself cannot explain the rapid rotation we see in the four other objects. It is possible that the magnetic field of GD 175 is truly a fossil field but we cannot rule out crystallization either.

\section{Acknowledgements}
This work is supported in part by the NSF under grants  AST-1906379 and AST-2205736, the NASA under grant 80NSSC22K0479, the NSERC Canada,
the Fund FRQ-NT (Qu\'ebec), and by the Smithsonian Institution. Based on observations obtained at the international Gemini Observatory, a program of NSF’s NOIRLab, which is managed by the Association of Universities for Research in Astronomy (AURA) under a cooperative agreement with the National Science Foundation on behalf of the Gemini Observatory partnership: the National Science Foundation (United States), National Research Council (Canada), Agencia Nacional de Investigación y Desarrollo (Chile), Ministerio de Ciencia, Tecnología e Innovación (Argentina), Ministério da Ciência, Tecnologia, Inovações e Comunicações (Brazil), and Korea Astronomy and Space Science Institute (Republic of Korea). This paper includes data gathered with the 6.5 meter Magellan Telescopes located at Las Campanas Observatory, Chile.

\section*{Data availability}

The data underlying this article are available in the Gemini Observatory Archive at https://archive.gemini.edu, and can be accessed with the program numbers GS-2019B-FT-107, GS-2020A-Q-311, GS-2021A-Q-136, GS-2021A-Q-321, GN-2020A-Q-116, GN-2021A-Q-135, and GN-2021A-Q-318.

\newcommand{\newblock}{}
\bibliographystyle{mnras}
\bibliography{MossMWD.bib}

\begin{thebibliography}{}
\makeatletter
\relax
\def\mn@urlcharsother{\let\do\@makeother \do\$\do\&\do\#\do\^\do\_\do\%\do\~}
\def\mn@doi{\begingroup\mn@urlcharsother \@ifnextchar [ {\mn@doi@}
  {\mn@doi@[]}}
\def\mn@doi@[#1]#2{\def\@tempa{#1}\ifx\@tempa\@empty \href
  {http://dx.doi.org/#2} {doi:#2}\else \href {http://dx.doi.org/#2} {#1}\fi
  \endgroup}
\def\mn@eprint#1#2{\mn@eprint@#1:#2::\@nil}
\def\mn@eprint@arXiv#1{\href {http://arxiv.org/abs/#1} {{\tt arXiv:#1}}}
\def\mn@eprint@dblp#1{\href {http://dblp.uni-trier.de/rec/bibtex/#1.xml}
  {dblp:#1}}
\def\mn@eprint@#1:#2:#3:#4\@nil{\def\@tempa {#1}\def\@tempb {#2}\def\@tempc
  {#3}\ifx \@tempc \@empty \let \@tempc \@tempb \let \@tempb \@tempa \fi \ifx
  \@tempb \@empty \def\@tempb {arXiv}\fi \@ifundefined
  {mn@eprint@\@tempb}{\@tempb:\@tempc}{\expandafter \expandafter \csname
  mn@eprint@\@tempb\endcsname \expandafter{\@tempc}}}

\bibitem[\protect\citeauthoryear{{Achilleos} \& {Wickramasinghe}}{{Achilleos}
  \& {Wickramasinghe}}{1989}]{Achilleos89}
{Achilleos} N.,  {Wickramasinghe} D.~T.,  1989, \mn@doi [\apj]
  {10.1086/168024}, \href
  {https://ui.adsabs.harvard.edu/abs/1989ApJ...346..444A} {346, 444}

\bibitem[\protect\citeauthoryear{{Amorim}, {Kepler}, {K{\"u}lebi}, {Jordan}  \&
  {Romero}}{{Amorim} et~al.}{2023}]{Amorim23}
{Amorim} L.~L.,  {Kepler} S.~O.,  {K{\"u}lebi} B.,  {Jordan} S.,   {Romero}
  A.~D.,  2023, \mn@doi [\apj] {10.3847/1538-4357/acaf6e}, \href
  {https://ui.adsabs.harvard.edu/abs/2023ApJ...944...56A} {944, 56}

\bibitem[\protect\citeauthoryear{{Bagnulo} \& {Landstreet}}{{Bagnulo} \&
  {Landstreet}}{2022}]{Bagnulo22}
{Bagnulo} S.,  {Landstreet} J.~D.,  2022, \mn@doi [\apjl]
  {10.3847/2041-8213/ac84d3}, \href
  {https://ui.adsabs.harvard.edu/abs/2022ApJ...935L..12B} {935, L12}

\bibitem[\protect\citeauthoryear{{B{\'e}dard}, {Bergeron}, {Brassard}  \&
  {Fontaine}}{{B{\'e}dard} et~al.}{2020}]{Bedard20}
{B{\'e}dard} A.,  {Bergeron} P.,  {Brassard} P.,   {Fontaine} G.,  2020,
  \mn@doi [\apj] {10.3847/1538-4357/abafbe}, \href
  {https://ui.adsabs.harvard.edu/abs/2020ApJ...901...93B} {901, 93}

\bibitem[\protect\citeauthoryear{{Bergeron}, {Ruiz}  \& {Leggett}}{{Bergeron}
  et~al.}{1992}]{Bergeron92}
{Bergeron} P.,  {Ruiz} M.-T.,   {Leggett} S.~K.,  1992, \mn@doi [\apj]
  {10.1086/171997}, \href
  {https://ui.adsabs.harvard.edu/abs/1992ApJ...400..315B} {400, 315}

\bibitem[\protect\citeauthoryear{{Bergeron}, {Ruiz}  \& {Leggett}}{{Bergeron}
  et~al.}{1997}]{Bergeron97}
{Bergeron} P.,  {Ruiz} M.~T.,   {Leggett} S.~K.,  1997, \mn@doi [\apjs]
  {10.1086/312955}, \href
  {https://ui.adsabs.harvard.edu/abs/1997ApJS..108..339B} {108, 339}

\bibitem[\protect\citeauthoryear{{Briggs}, {Ferrario}, {Tout}, {Wickramasinghe}
   \& {Hurley}}{{Briggs} et~al.}{2015}]{Briggs15}
{Briggs} G.~P.,  {Ferrario} L.,  {Tout} C.~A.,  {Wickramasinghe} D.~T.,
  {Hurley} J.~R.,  2015, \mn@doi [\mnras] {10.1093/mnras/stu2539}, \href
  {https://ui.adsabs.harvard.edu/abs/2015MNRAS.447.1713B} {447, 1713}

\bibitem[\protect\citeauthoryear{{Briggs}, {Ferrario}, {Tout}  \&
  {Wickramasinghe}}{{Briggs} et~al.}{2018a}]{Briggs18}
{Briggs} G.~P.,  {Ferrario} L.,  {Tout} C.~A.,   {Wickramasinghe} D.~T.,
  2018a, \mn@doi [\mnras] {10.1093/mnras/sty1150}, \href
  {https://ui.adsabs.harvard.edu/abs/2018MNRAS.478..899B} {478, 899}

\bibitem[\protect\citeauthoryear{{Briggs}, {Ferrario}, {Tout}  \&
  {Wickramasinghe}}{{Briggs} et~al.}{2018b}]{Briggs18b}
{Briggs} G.~P.,  {Ferrario} L.,  {Tout} C.~A.,   {Wickramasinghe} D.~T.,
  2018b, \mn@doi [\mnras] {10.1093/mnras/sty2481}, \href
  {https://ui.adsabs.harvard.edu/abs/2018MNRAS.481.3604B} {481, 3604}

\bibitem[\protect\citeauthoryear{{Caron}, {Bergeron}, {Blouin}  \&
  {Leggett}}{{Caron} et~al.}{2023}]{Caron23}
{Caron} A.,  {Bergeron} P.,  {Blouin} S.,   {Leggett} S.~K.,  2023, \mn@doi
  [\mnras] {10.1093/mnras/stac3733}, \href
  {https://ui.adsabs.harvard.edu/abs/2023MNRAS.519.4529C} {519, 4529}

\bibitem[\protect\citeauthoryear{{Coutu}, {Dufour}, {Bergeron}, {Blouin},
  {Loranger}, {Allard}  \& {Dunlap}}{{Coutu} et~al.}{2019}]{Coutu19}
{Coutu} S.,  {Dufour} P.,  {Bergeron} P.,  {Blouin} S.,  {Loranger} E.,
  {Allard} N.~F.,   {Dunlap} B.~H.,  2019, \mn@doi [\apj]
  {10.3847/1538-4357/ab46b9}, \href
  {https://ui.adsabs.harvard.edu/abs/2019ApJ...885...74C} {885, 74}

\bibitem[\protect\citeauthoryear{{Cummings}, {Kalirai}, {Tremblay},
  {Ramirez-Ruiz}  \& {Choi}}{{Cummings} et~al.}{2018}]{Cummings18}
{Cummings} J.~D.,  {Kalirai} J.~S.,  {Tremblay} P.~E.,  {Ramirez-Ruiz} E.,
  {Choi} J.,  2018, \mn@doi [\apj] {10.3847/1538-4357/aadfd6}, \href
  {https://ui.adsabs.harvard.edu/abs/2018ApJ...866...21C} {866, 21}

\bibitem[\protect\citeauthoryear{{De Lee} et~al.,}{{De Lee}
  et~al.}{2013}]{DeLee13}
{De Lee} N.,  et~al., 2013, \mn@doi [\aj] {10.1088/0004-6256/145/6/155}, \href
  {https://ui.adsabs.harvard.edu/abs/2013AJ....145..155D} {145, 155}

\bibitem[\protect\citeauthoryear{{Dunlap} \& {Clemens}}{{Dunlap} \&
  {Clemens}}{2015}]{Dunlap15}
{Dunlap} B.~H.,  {Clemens} J.~C.,  2015, in {Dufour} P.,  {Bergeron} P.,
  {Fontaine} G.,  eds,  Astronomical Society of the Pacific Conference Series
  Vol. 493, 19th European Workshop on White Dwarfs. p.~547

\bibitem[\protect\citeauthoryear{{Ferrario}, {de Martino}  \&
  {G{\"a}nsicke}}{{Ferrario} et~al.}{2015}]{Ferrario15a}
{Ferrario} L.,  {de Martino} D.,   {G{\"a}nsicke} B.~T.,  2015, \mn@doi [\ssr]
  {10.1007/s11214-015-0152-0}, \href
  {https://ui.adsabs.harvard.edu/abs/2015SSRv..191..111F} {191, 111}

\bibitem[\protect\citeauthoryear{{Gentile Fusillo}, {Tremblay}, {Jordan},
  {G{\"a}nsicke}, {Kalirai}  \& {Cummings}}{{Gentile Fusillo}
  et~al.}{2018}]{Gentile18}
{Gentile Fusillo} N.~P.,  {Tremblay} P.~E.,  {Jordan} S.,  {G{\"a}nsicke}
  B.~T.,  {Kalirai} J.~S.,   {Cummings} J.,  2018, \mn@doi [\mnras]
  {10.1093/mnras/stx2584}, \href
  {https://ui.adsabs.harvard.edu/abs/2018MNRAS.473.3693G} {473, 3693}

\bibitem[\protect\citeauthoryear{{Gentile Fusillo} et~al.,}{{Gentile Fusillo}
  et~al.}{2021}]{Gentile21}
{Gentile Fusillo} N.~P.,  et~al., 2021, \mn@doi [\mnras]
  {10.1093/mnras/stab2672}, \href
  {https://ui.adsabs.harvard.edu/abs/2021MNRAS.508.3877G} {508, 3877}

\bibitem[\protect\citeauthoryear{{Giammichele} et~al.,}{{Giammichele}
  et~al.}{2018}]{Giammichele18}
{Giammichele} N.,  et~al., 2018, \mn@doi [\nat] {10.1038/nature25136}, \href
  {https://ui.adsabs.harvard.edu/abs/2018Natur.554...73G} {554, 73}

\bibitem[\protect\citeauthoryear{{Giammichele}, {Charpinet}  \&
  {Brassard}}{{Giammichele} et~al.}{2022}]{Giammichele22}
{Giammichele} N.,  {Charpinet} S.,   {Brassard} P.,  2022, \mn@doi [Frontiers
  in Astronomy and Space Sciences] {10.3389/fspas.2022.879045}, \href
  {https://ui.adsabs.harvard.edu/abs/2022FrASS...9.9045G} {9, 879045}

\bibitem[\protect\citeauthoryear{{Ginzburg}, {Fuller}, {Kawka}  \&
  {Caiazzo}}{{Ginzburg} et~al.}{2022}]{Ginzburg22}
{Ginzburg} S.,  {Fuller} J.,  {Kawka} A.,   {Caiazzo} I.,  2022, \mn@doi
  [\mnras] {10.1093/mnras/stac1363}, \href
  {https://ui.adsabs.harvard.edu/abs/2022MNRAS.514.4111G} {514, 4111}

\bibitem[\protect\citeauthoryear{{Hardy}, {Dufour}  \& {Jordan}}{{Hardy}
  et~al.}{2023}]{Hardy23}
{Hardy} F.,  {Dufour} P.,   {Jordan} S.,  2023, \mn@doi [\mnras]
  {10.1093/mnras/stad196}, \href
  {https://ui.adsabs.harvard.edu/abs/2023MNRAS.520.6111H} {520, 6111}

\bibitem[\protect\citeauthoryear{{Hermes} et~al.,}{{Hermes}
  et~al.}{2017}]{Hermes17}
{Hermes} J.~J.,  et~al., 2017, \mn@doi [\apjs] {10.3847/1538-4365/aa8bb5},
  \href {https://ui.adsabs.harvard.edu/abs/2017ApJS..232...23H} {232, 23}

\bibitem[\protect\citeauthoryear{{Isern}, {Garc{\'\i}a-Berro}, {K{\"u}lebi}  \&
  {Lor{\'e}n-Aguilar}}{{Isern} et~al.}{2017}]{Isern17}
{Isern} J.,  {Garc{\'\i}a-Berro} E.,  {K{\"u}lebi} B.,   {Lor{\'e}n-Aguilar}
  P.,  2017, \mn@doi [\apjl] {10.3847/2041-8213/aa5eae}, \href
  {https://ui.adsabs.harvard.edu/abs/2017ApJ...836L..28I} {836, L28}

\bibitem[\protect\citeauthoryear{{Kawka}, {Briggs}, {Vennes}, {Ferrario},
  {Paunzen}  \& {Wickramasinghe}}{{Kawka} et~al.}{2017}]{Kawka17}
{Kawka} A.,  {Briggs} G.~P.,  {Vennes} S.,  {Ferrario} L.,  {Paunzen} E.,
  {Wickramasinghe} D.~T.,  2017, \mn@doi [\mnras] {10.1093/mnras/stw3149},
  \href {https://ui.adsabs.harvard.edu/abs/2017MNRAS.466.1127K} {466, 1127}

\bibitem[\protect\citeauthoryear{{Kawka}, {Ferrario}  \& {Vennes}}{{Kawka}
  et~al.}{2023}]{Kawka23}
{Kawka} A.,  {Ferrario} L.,   {Vennes} S.,  2023, \mn@doi [\mnras]
  {10.1093/mnras/stad553}, \href
  {https://ui.adsabs.harvard.edu/abs/2023MNRAS.520.6299K} {520, 6299}

\bibitem[\protect\citeauthoryear{{Kilic}, {Rolland}, {Bergeron}, {Vanderbosch},
  {Benni}  \& {Garlitz}}{{Kilic} et~al.}{2019}]{Kilic19}
{Kilic} M.,  {Rolland} B.,  {Bergeron} P.,  {Vanderbosch} Z.,  {Benni} P.,
  {Garlitz} J.,  2019, \mn@doi [\mnras] {10.1093/mnras/stz2394}, \href
  {https://ui.adsabs.harvard.edu/abs/2019MNRAS.489.3648K} {489, 3648}

\bibitem[\protect\citeauthoryear{{Kilic}, {Bergeron}, {Kosakowski}, {Brown},
  {Ag{\"u}eros}  \& {Blouin}}{{Kilic} et~al.}{2020}]{Kilic20}
{Kilic} M.,  {Bergeron} P.,  {Kosakowski} A.,  {Brown} W.~R.,  {Ag{\"u}eros}
  M.~A.,   {Blouin} S.,  2020, \mn@doi [\apj] {10.3847/1538-4357/ab9b8d}, \href
  {https://ui.adsabs.harvard.edu/abs/2020ApJ...898...84K} {898, 84}

\bibitem[\protect\citeauthoryear{{Kilic} et~al.,}{{Kilic}
  et~al.}{2023}]{Kilic23}
{Kilic} M.,  et~al., 2023, \mn@doi [\mnras] {10.1093/mnras/stac3182}, \href
  {https://ui.adsabs.harvard.edu/abs/2023MNRAS.518.2341K} {518, 2341}

\bibitem[\protect\citeauthoryear{{Landstreet} \& {Bagnulo}}{{Landstreet} \&
  {Bagnulo}}{2020}]{Landstreet20}
{Landstreet} J.~D.,  {Bagnulo} S.,  2020, \mn@doi [\aap]
  {10.1051/0004-6361/201937301}, \href
  {https://ui.adsabs.harvard.edu/abs/2020A&A...634L..10L} {634, L10}

\bibitem[\protect\citeauthoryear{{Liebert} et~al.,}{{Liebert}
  et~al.}{2005}]{Liebert05}
{Liebert} J.,  et~al., 2005, \mn@doi [\aj] {10.1086/429639}, \href
  {https://ui.adsabs.harvard.edu/abs/2005AJ....129.2376L} {129, 2376}

\bibitem[\protect\citeauthoryear{{Liebert}, {Ferrario}, {Wickramasinghe}  \&
  {Smith}}{{Liebert} et~al.}{2015}]{Liebert15}
{Liebert} J.,  {Ferrario} L.,  {Wickramasinghe} D.~T.,   {Smith} P.~S.,  2015,
  \mn@doi [\apj] {10.1088/0004-637X/804/2/93}, \href
  {https://ui.adsabs.harvard.edu/abs/2015ApJ...804...93L} {804, 93}

\bibitem[\protect\citeauthoryear{{Manser} et~al.,}{{Manser}
  et~al.}{2023}]{Manser23}
{Manser} C.~J.,  et~al., 2023, \mn@doi [\mnras] {10.1093/mnras/stad727}, \href
  {https://ui.adsabs.harvard.edu/abs/2023MNRAS.521.4976M} {521, 4976}

\bibitem[\protect\citeauthoryear{{Parsons}, {G{\"a}nsicke}, {Schreiber},
  {Marsh}, {Ashley}, {Breedt}, {Littlefair}  \& {Meusinger}}{{Parsons}
  et~al.}{2021}]{Parsons21}
{Parsons} S.~G.,  {G{\"a}nsicke} B.~T.,  {Schreiber} M.~R.,  {Marsh} T.~R.,
  {Ashley} R.~P.,  {Breedt} E.,  {Littlefair} S.~P.,   {Meusinger} H.,  2021,
  \mn@doi [\mnras] {10.1093/mnras/stab284}, \href
  {https://ui.adsabs.harvard.edu/abs/2021MNRAS.502.4305P} {502, 4305}

\bibitem[\protect\citeauthoryear{{Pereira}, {Bergeron}  \&
  {Wesemael}}{{Pereira} et~al.}{2005}]{Pereira05}
{Pereira} C.,  {Bergeron} P.,   {Wesemael} F.,  2005, \mn@doi [\apj]
  {10.1086/429219}, \href
  {https://ui.adsabs.harvard.edu/abs/2005ApJ...623.1076P} {623, 1076}

\bibitem[\protect\citeauthoryear{{Reding}, {Hermes}, {Clemens}, {Hegedus}  \&
  {Kaiser}}{{Reding} et~al.}{2023}]{Reding23}
{Reding} J.~S.,  {Hermes} J.~J.,  {Clemens} J.~C.,  {Hegedus} R.~J.,   {Kaiser}
  B.~C.,  2023, \mn@doi [\mnras] {10.1093/mnras/stad760}, \href
  {https://ui.adsabs.harvard.edu/abs/2023MNRAS.tmp..782R} {}

\bibitem[\protect\citeauthoryear{{Rolland} \& {Bergeron}}{{Rolland} \&
  {Bergeron}}{2015}]{Rolland15}
{Rolland} B.,  {Bergeron} P.,  2015, in {Dufour} P.,  {Bergeron} P.,
  {Fontaine} G.,  eds,  Astronomical Society of the Pacific Conference Series
  Vol. 493, 19th European Workshop on White Dwarfs. p.~53

\bibitem[\protect\citeauthoryear{{Ruiter}}{{Ruiter}}{2020}]{Ruiter20}
{Ruiter} A.~J.,  2020, \mn@doi [IAU Symposium] {10.1017/S1743921320000587},
  \href {https://ui.adsabs.harvard.edu/abs/2020IAUS..357....1R} {357, 1}

\bibitem[\protect\citeauthoryear{{Schreiber}, {Belloni}, {G{\"a}nsicke},
  {Parsons}  \& {Zorotovic}}{{Schreiber} et~al.}{2021}]{Schreiber21}
{Schreiber} M.~R.,  {Belloni} D.,  {G{\"a}nsicke} B.~T.,  {Parsons} S.~G.,
  {Zorotovic} M.,  2021, \mn@doi [Nature Astronomy]
  {10.1038/s41550-021-01346-8}, \href
  {https://ui.adsabs.harvard.edu/abs/2021NatAs...5..648S} {5, 648}

\bibitem[\protect\citeauthoryear{{Subasavage}, {Henry}, {Bergeron}, {Dufour},
  {Hambly}  \& {Beaulieu}}{{Subasavage} et~al.}{2007}]{Subasavage07}
{Subasavage} J.~P.,  {Henry} T.~J.,  {Bergeron} P.,  {Dufour} P.,  {Hambly}
  N.~C.,   {Beaulieu} T.~D.,  2007, in {Napiwotzki} R.,  {Burleigh} M.~R.,
  eds,  Astronomical Society of the Pacific Conference Series Vol. 372, 15th
  European Workshop on White Dwarfs. p.~53 (\mn@eprint {arXiv}
  {astro-ph/0610946}), \mn@doi{10.48550/arXiv.astro-ph/0610946}

\bibitem[\protect\citeauthoryear{{Temmink}, {Toonen}, {Zapartas}, {Justham}  \&
  {G{\"a}nsicke}}{{Temmink} et~al.}{2020}]{Temmink20}
{Temmink} K.~D.,  {Toonen} S.,  {Zapartas} E.,  {Justham} S.,   {G{\"a}nsicke}
  B.~T.,  2020, \mn@doi [\aap] {10.1051/0004-6361/201936889}, \href
  {https://ui.adsabs.harvard.edu/abs/2020A&A...636A..31T} {636, A31}

\bibitem[\protect\citeauthoryear{{Tout}, {Wickramasinghe}  \&
  {Ferrario}}{{Tout} et~al.}{2004}]{Tout04}
{Tout} C.~A.,  {Wickramasinghe} D.~T.,   {Ferrario} L.,  2004, \mn@doi [\mnras]
  {10.1111/j.1365-2966.2004.08482.x}, \href
  {https://ui.adsabs.harvard.edu/abs/2004MNRAS.355L..13T} {355, L13}

\bibitem[\protect\citeauthoryear{{Tout}, {Wickramasinghe}, {Liebert},
  {Ferrario}  \& {Pringle}}{{Tout} et~al.}{2008}]{Tout08}
{Tout} C.~A.,  {Wickramasinghe} D.~T.,  {Liebert} J.,  {Ferrario} L.,
  {Pringle} J.~E.,  2008, \mn@doi [\mnras] {10.1111/j.1365-2966.2008.13291.x},
  \href {https://ui.adsabs.harvard.edu/abs/2008MNRAS.387..897T} {387, 897}

\bibitem[\protect\citeauthoryear{{Tremblay}, {Fontaine}, {Freytag}, {Steiner},
  {Ludwig}, {Steffen}, {Wedemeyer}  \& {Brassard}}{{Tremblay}
  et~al.}{2015}]{Tremblay15}
{Tremblay} P.~E.,  {Fontaine} G.,  {Freytag} B.,  {Steiner} O.,  {Ludwig}
  H.~G.,  {Steffen} M.,  {Wedemeyer} S.,   {Brassard} P.,  2015, \mn@doi [\apj]
  {10.1088/0004-637X/812/1/19}, \href
  {https://ui.adsabs.harvard.edu/abs/2015ApJ...812...19T} {812, 19}

\bibitem[\protect\citeauthoryear{{Wickramasinghe} \&
  {Ferrario}}{{Wickramasinghe} \& {Ferrario}}{2005}]{Wick05}
{Wickramasinghe} D.~T.,  {Ferrario} L.,  2005, \mn@doi [\mnras]
  {10.1111/j.1365-2966.2004.08603.x}, \href
  {https://ui.adsabs.harvard.edu/abs/2005MNRAS.356.1576W} {356, 1576}

\makeatother
\end{thebibliography}

\appendix

\section{Nonvariable Targets \label{app}}

Here we present the trailed spectra and the best-fitting magnetic models to three of our targets that did not show significant variations in their spectra.

\begin{figure*}
 \centering
 \includegraphics[width=2.3in,clip=true,trim=7in 0.7in 6.5in 0in]{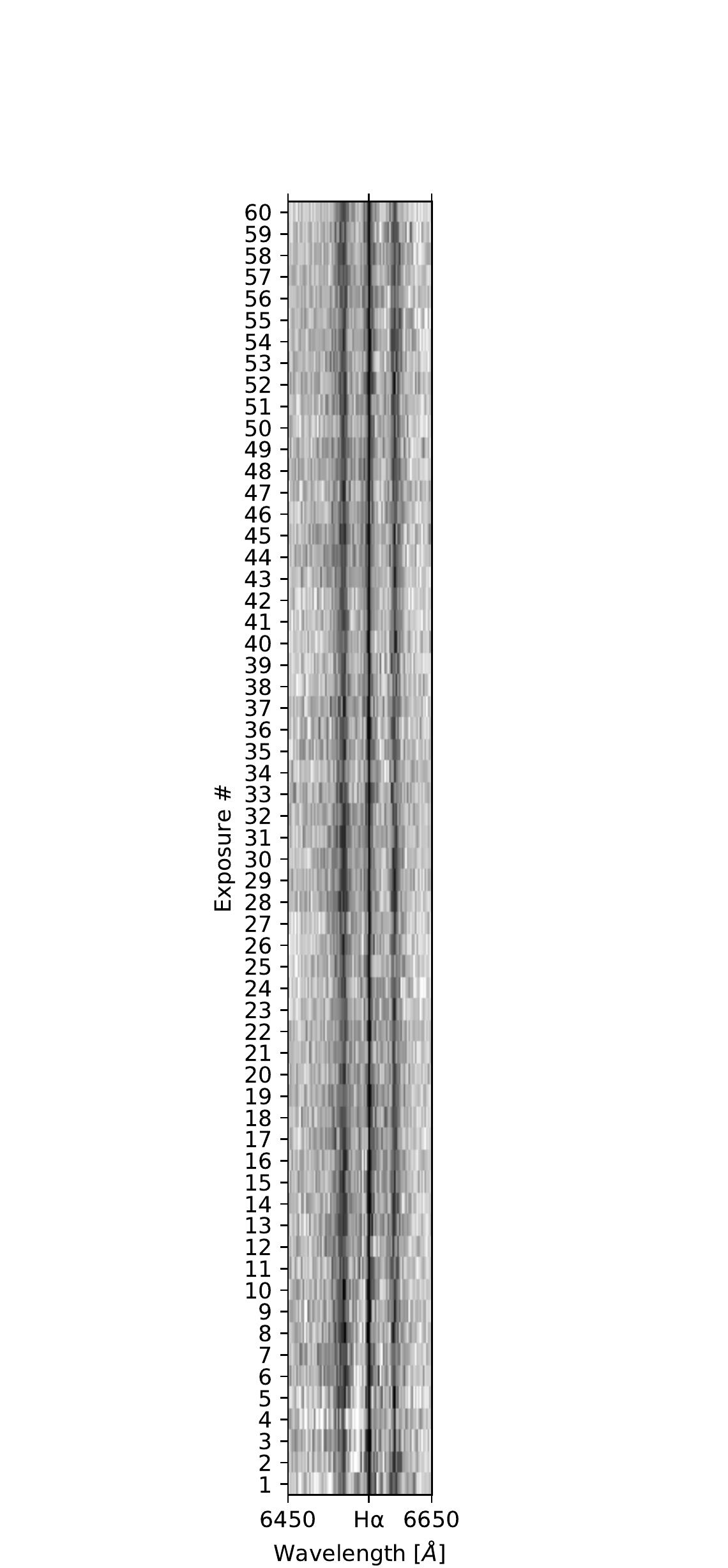}
 \includegraphics[width=1.5in,clip=true,trim=2.3in 0.05in 2.3in 0in]{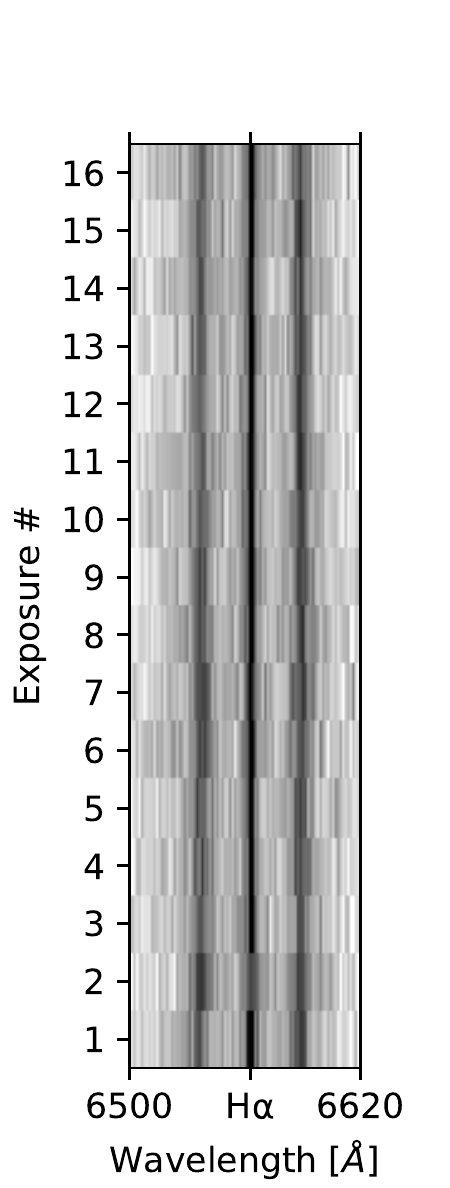}
 \includegraphics[width=2.3in,clip=true,trim=3.5in 0.25in 3.5in 0in]{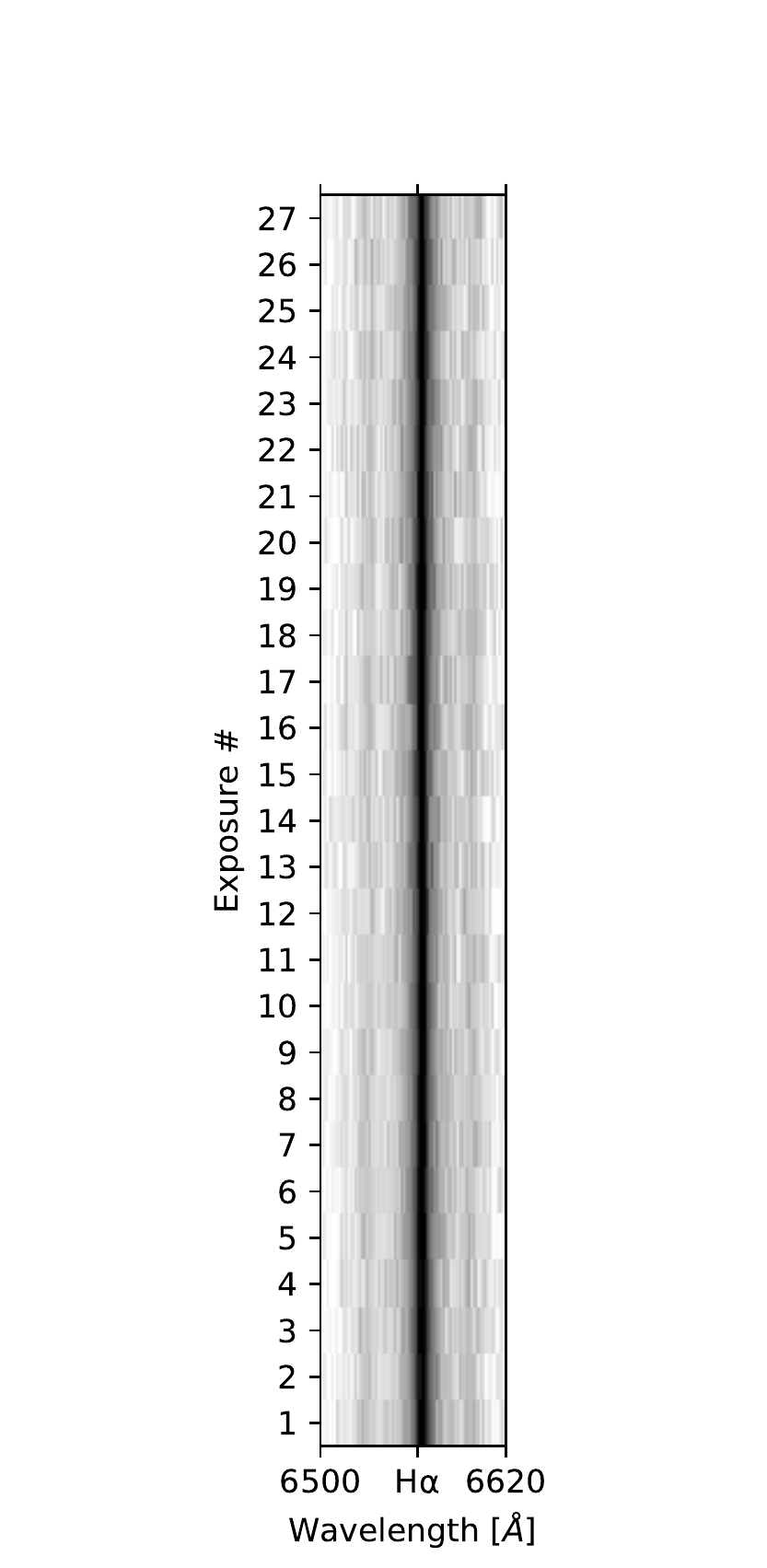}
 \caption{Trailed spectra for our remaining targets that do not show significant spectroscopic variations: GD 175, LP 226-48, and G160-51.}
 \label{12}
\end{figure*}

\begin{figure*}
 \centering
 \includegraphics[width=3.2in,clip=true,trim=1.8in 1.35in 2in 1.5in]{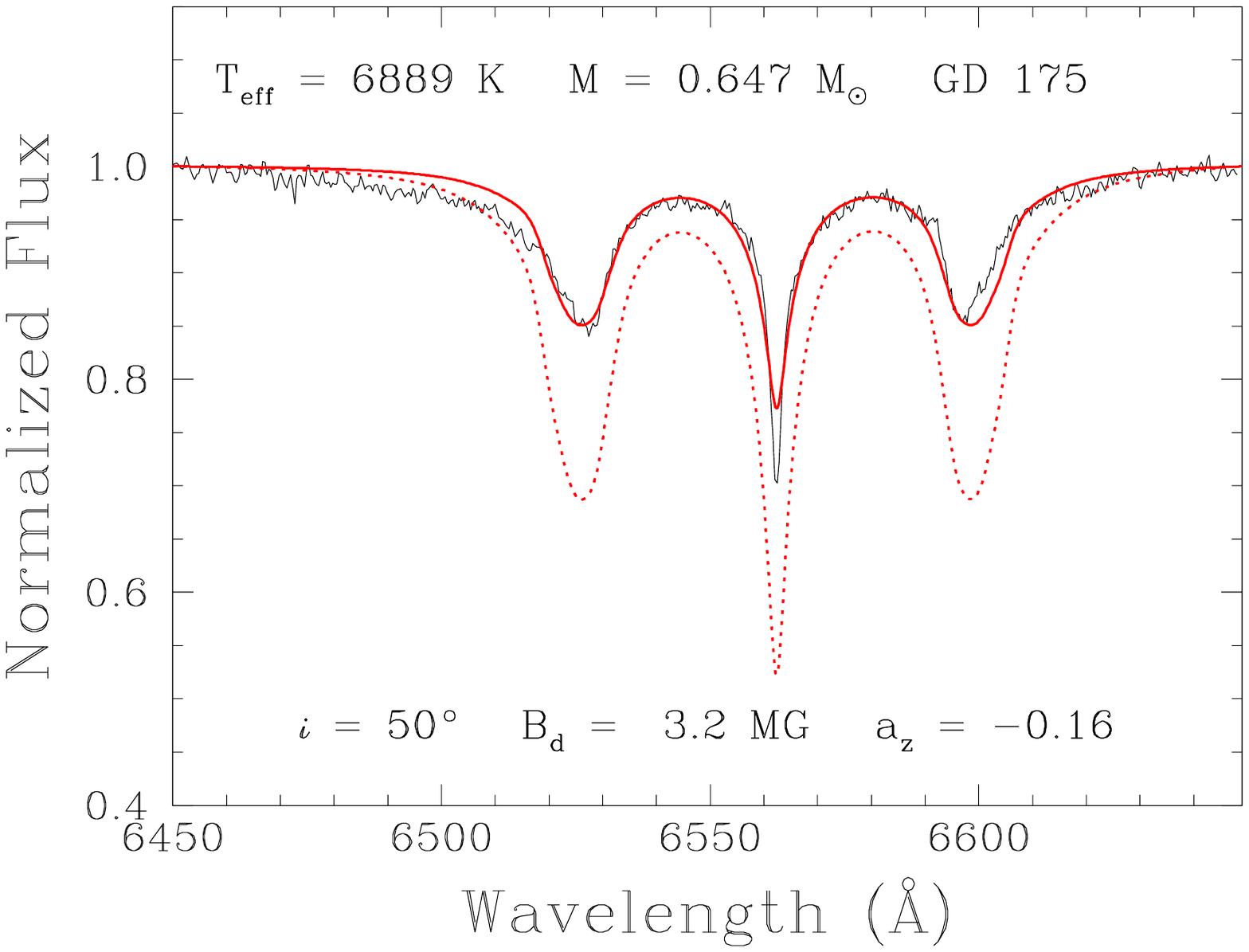}
 \includegraphics[width=3.2in,clip=true,trim=1.8in 1.35in 2in 1.5in]{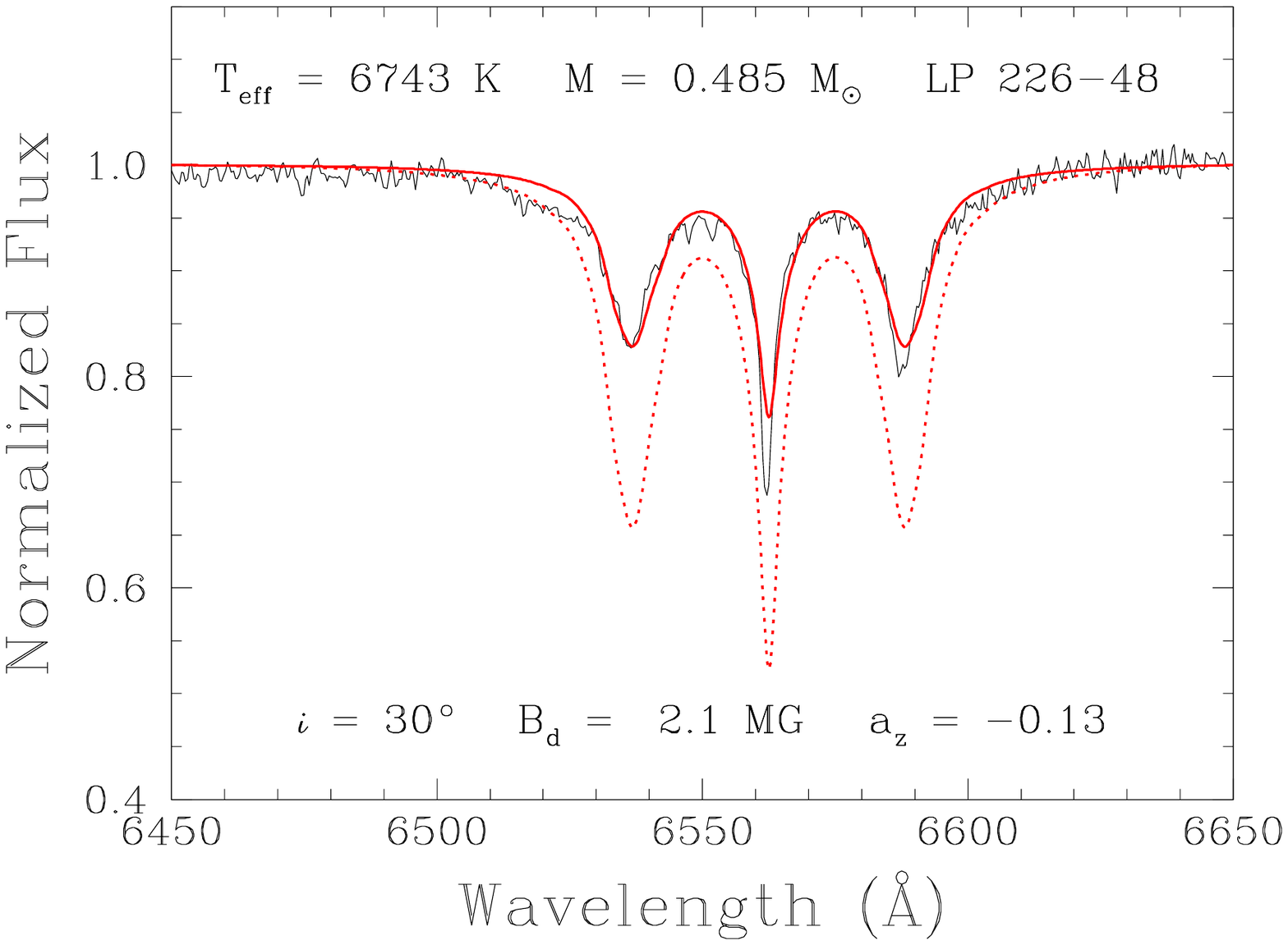}
 \includegraphics[width=3.2in,clip=true,trim=1.8in 1.35in 2in 1.5in]{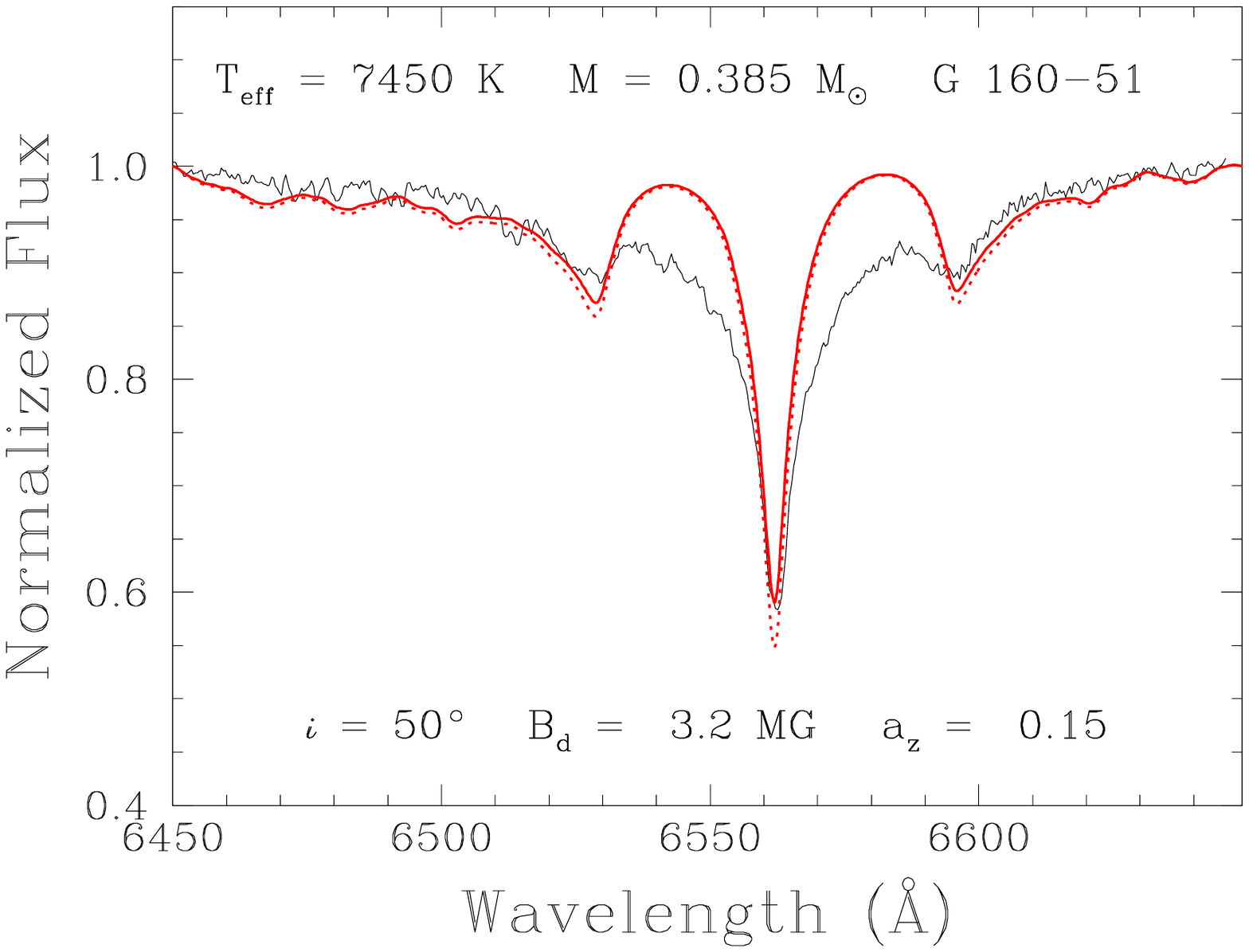}
 \caption{Fits for the combined spectra of GD 175, LP 226-48, and G160-51. The poor fit for G160-51 (bottom) suggests this target may be in a binary with a DA companion.}
 \label{13}
\end{figure*}

\end{document}